\makeatletter
\declare@file@substitution{revtex4-1.cls}{revtex4-2.cls}
\makeatother

\documentclass{aastex631}

\usepackage{savesym}
\savesymbol{tablenum}

\usepackage{siunitx} 
\usepackage{xcolor}

\restoresymbol{SIX}{tablenum}

\shorttitle{MIRI source number counts and properties}
\shortauthors{Sajkov et al.}

\begin{document}

\title{Halfway to the Peak: The {\sl JWST} MIRI 5.6 micron number counts and source population}
\author{Leonid Sajkov}
\affiliation{Tufts University, Medford, MA}
\author{Anna Sajina}
\affiliation{Tufts University, Medford, MA}
\author{Alexandra Pope}
\affiliation{University of Massachusetts at Amherst, Amherst, MA}
\author{Stacey Alberts}
\affiliation{Steward Observatory, University of Arizona, Tucson, AZ}
\author{Lee Armus}
\affiliation{California Institute of Technology, Pasadena, CA}
\author{Duncan Farrah}
\affiliation{University of Hawaii at Manoa, Honolulu, HI}
\author{Jamie Lin}
\affiliation{Tufts University, Medford, MA}
\author{Danilo Marchesini}
\affiliation{Tufts University, Medford, MA}
\author{Jed McKinney}
\affiliation{University of  Texas at Austin, Austin, TX}
\author{Sylvain Veilleux}
\affiliation{University of Maryland, College Park, MD}
\author{Lin Yan}
\affiliation{California Institute of Technology, Pasadena, CA}
\author{Jason Young}
\affiliation{Williams College, Williamstown, MA}
\affiliation{SETI Institute, Mountain View, CA}

\begin{abstract}
We present an analysis of eight JWST Mid-Infrared Instrument (MIRI) 5.6 micron images with $5\,\sigma$ depths of $\approx$\SI{0.1}{\micro Jy}. We detect 2854 sources within our combined area of 18.4\,square arcminutes. We compute the MIRI \SI{5.6}{\micro\meter} number counts including an analysis of the field-to-field variation.  Compared to earlier published MIRI \SI{5.6}{\micro\meter} counts, our counts have a more pronounced knee, at roughly \SI{2}{\micro Jy}. The location and amplitude of the counts at the knee are consistent with the Cowley et al. (2018) model predictions, although these models tend to overpredict the counts below the knee. In areas of overlap, 84\,\% of the MIRI sources have a counterpart in the COSMOS2020 catalog. These MIRI sources have redshifts that are mostly in the $z\sim0.5-2$, with a tail out to $z\sim5$. They are predominantly moderate to low stellar masses ($10^8-10^{10}$M$_{\odot}$) main sequence star-forming galaxies{\bf ,} suggesting that with $\approx$2\,hr exposures, MIRI can reach well below $M^*$ at cosmic noon and reach higher mass systems out to $z\sim5$. Nearly 70\% of the COSMOS2020 sources in areas of overlap now have a data point at \SI{5.6}{\micro \meter} (rest-frame near-IR at cosmic noon) which allows for more accurate stellar population parameter estimates. Finally, we discover 31 MIRI-bright sources not present in COSMOS2020. A cross-match with IRAC channel 1 suggests that 10-20\,\% of these are likely lower mass (M$_*\approx10^9$M$_{\odot}$), $z\sim1$ dusty galaxies. The rest (80--90\%) are consistent with more massive but still very dusty galaxies at $z>3$.
\end{abstract}

\section{Introduction}

Extragalactic astronomy aims to study how galaxies form and evolve across time. To do so, we build multiwavelength galaxy surveys which allow us to compare the data with spectral energy distribution (SED) models to infer redshifts and stellar population parameters \citep[e.g.][]{Brammer2008,Weaver2022,Wang2024}. Mid-IR data are particularly critical to this effort since they probe the rest-frame near-IR in the cosmic noon epoch and earlier which allows for much more accurate stellar masses and star formation rates (SFRs) \citep[e.g.][]{Elsner2008,Stefanon2015,Muzzin2009,martis2023,LaTorre2024}. Indeed, over the past two decades {\sl Spitzer} IRAC data -- especially its first two channels, available even in the warm mission \citep{Lacy2021,Annunziatella2023} have been a critical component of these multiwavelength surveys. With IRAC channel 1 and 2, we reached stellar masses of $\approx10^{9.5}$M$\odot$ at cosmic noon, i.e. well below the knee of the SMF \citep{Elsner2008, MadauDickinson2014}, but only detected the most massive galaxies ($log(M_*)>11$) at $z>4$ \citep[e.g.][]{Stefanon2015}. The IRAC channel 3, at \SI{5.8}{\micro \meter}, which critically covers the rest-frame of the stellar \SI{1.6}{\micro \meter} bump at $z\sim2-3$ was relatively underutilized because it was not available during the extended warm mission and was the least sensitive IRAC channel even during the cold mission.  

The \textit{JWST} \citep{Gardner2006}, which launched on December 25, 2021, is already revolutionizing infrared astronomy with its unprecedented performance \citep{rigby:2023}. In particular, the Mid-InfraRed Instrument \citep[MIRI;][]{Rieke2015} allows for significantly greater sensitivity and angular resolution relative to the {\sl Spitzer} Space Telescope. Recently published {\sl JWST}/MIRI number counts at \SI{7.7}{\micro \meter}, \SI{10}{\micro \meter} and \SI{15}{\micro \meter} bands \citep{Ling2022,wu:2023,Kirkpatrick2023,Stone2024} show dramatic improvement in the depth reached, even with moderate exposure times, relative to prior measurements from {\sl Spitzer} and {\sl ISO}. These counts have already been used in constraining galaxy and black hole evolution models \citep{Kim2024}. Number counts at the shortest MIRI wavelength (\SI{5.6}{\micro \meter}) are much more scarse \citep{Yang2023,Stone2024}, but they are critical as this band samples the stellar \SI{1.6}{\micro \meter} bump at $z\sim2-3$ and thus is critical in testing our models of the galaxy population at cosmic noon.

In addition, sampling the rest-frame near-IR at cosmic noon, the \SI{5.6}{\micro \meter} band is much less sensitive to the effects of dust obscuration even than traditionally `dust clean' bands such as the $K$ band, which is rest-frame $r$-band at the same redshifts. This insensitivity to dust obscuration is important as in the more than two decades since the discovery of the Cosmic Infrared Background \citep[e.g.][]{Puget1996} it has become abundantly clear that the bulk of star-formation activity at cosmic noon and beyond takes place in dust obscured galaxies (see \citealt{Casey2014} for a review) which in their extreme are missed in UV/optical surveys \citep[see e.g.][among many others]{Hughes1998,Sajina2006,Pope2008,Zavala2021}. Recently, {\sl JWST} data have further highlighted this by finding that even deep {\sl HST} data miss the reddest/most dust obscured parts of the galaxy population \citep[e.g.][]{Labbe2023,Barrufet2023,Williams2023}. 

In this paper, we use deep MIRI \SI{5.6}{\micro \meter} images to provide a first look at the properties of the \SI{5.6}{\micro \meter} number counts and source population based on images obtained with nearly 2\,hr exposure times. This wavelength corresponds roughly to the rest-frame $H/K$-band at cosmic noon ($z\sim1-3$), thus probing primarily stellar mass at these critical redshifts. This paper is structured as follows. In Section\,\ref{sec:data} we present the MIRI \SI{5.6}{\micro \meter} imaging data and data reduction. In Section\,\ref{sec:analysis} we present the source detection and photometry and the verification of the latter. In Section\,\ref{sec:results}, we show the key results of our work. These include the \SI{5.6}{\micro \meter} number counts (Section\,\ref{subsec:number_counts}); the redshift distribution and other properties of the MIRI\,\SI{5.6}{\micro \meter} sources with counterparts in COSMOS2020 (Section\,\ref{subsec:properties}); and lastly MIRI \SI{5.6}{\micro \meter} sources without counterparts in COSMOS2020 suggesting very red colors (Section\,\ref{subsec:veryredmiri}). Throughout we adopt the AB magnitude system \citep{OkeGunn1974}. We adopt the cosmology model from the COSMOS2020 catalog \citep{Weaver2022}, which is a standard $\Lambda$CDM cosmology with H$_0$ = 70 km/s/Mpc, $\Omega_{M,0}$ = 0.3, and $\Omega_{\Lambda,0}$ = 0.7. 

\section{Data} 
\label{sec:data}

\subsection{MIRI imaging data}
Our data come from {\sl JWST}/MIRI imaging with the \SI{5.6}{\micro \meter} filter obtained as part of the GO1 program ``Halfway to the Peak: A Bridge Program To Map Coeval Star Formation and Supermassive Black Hole Growth" \citep[PIs Pope, Sajina, Yan; PID 1762;][]{halfway:2021}. This program observed eight targets with the MIRI medium-resolution spectrometer (MRS) instrument: details on the targets and MRS spectra analysis are provided in \citet{Young2023}.  Simultaneous with the MIRI/MRS observations, we obtained MIRI \SI{5.6}{\micro \meter} imaging in fields adjacent to the MRS targets. This parallel observing means our imaging fields are effectively blank fields and therefore ideal for statistical studies such as number counts. In total, we have six fields in the First Look Survey (FLS) field and two in the COSMOS field. The FLS fields were observed in July/August 2022 and the COSMOS fields were observed in December 2022. 

\begin{table}[h]
\centering
\begin{tabular}{cccc}
 \hline
 Field name & RA & Dec & Exp. time \\ [0.5ex] 
 \hline\hline
      FLS1 & 17:12:28.50 & +58:59:30.12 & 1.85hrs\\
      FLS2 & 17:24:46.96 & +59:15:24.01 & 1.85hrs\\
      FLS3 & 17:21:07.16 & +58:45:39.67 & 1.85hrs\\
      FLS4 & 17:22:49.98 & +59:40:32.35 & 1.85hrs\\
      FLS5 & 17:19:12.87 & +59:28:53.55 & 1.85hrs\\
      FLS6 & 17:13:13.14 & +58:55:22.51 & 1.85hrs\\
   COSMOS1 & 10:01:14.98 &  +2:24:36.36 & 1.85hrs\\
   COSMOS2 & 10:01:42.53 &  +2:47:26.64 & 1.85hrs\\
 \hline
 \end{tabular}
 \caption{The eight MIRI imaging fields used in this paper. The coordinates given are for the centers of the fields. Each field covers $\approx$2.42\,square arcminutes for a total of $\approx$19.4\,square arcminutes, which is reduced to 18.4\,square arcminutes after masking. }
 \label{table:observations}
\end{table}

\begin{figure}[h!]
    \centering
    \includegraphics[width=0.99\linewidth]{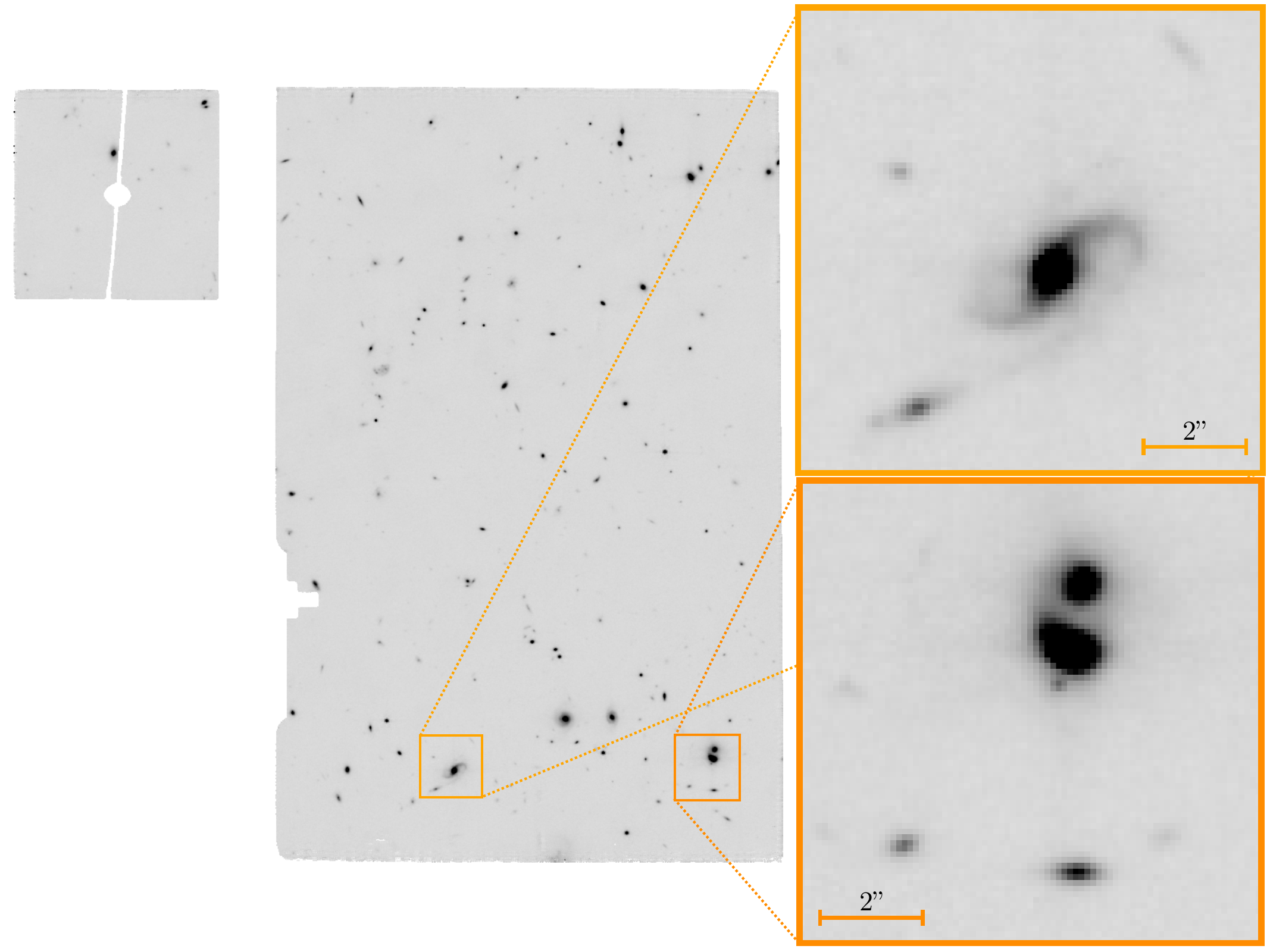}
    \caption{An example of our MIRI images (for the FLS6 field) highlighting the number of sources clearly detected within the imager's FOV of only 2.3\,sq.arcmin. The angular resolution is also illustrated in the zoomed-in insets, both measuring 9$\times$\ang{;;9}, where we see both resolved morphology as illustrated by the detected spiral structure (\textit{top inset}) as well as resolved closely spaced sources (that would have been blended in the {\sl Spitzer} IRAC images, \textit{middle inset}). The 2$^{\prime\prime}$ scale bars are comparable to the IRAC channel 3 FWHM which is 1.88$^{\prime\prime}$.}
    \label{fig:miri_images}
\end{figure}

We obtain data over the MIRI imager with an FOV of $74\times$\ang{;;113} as well as the smaller MIRI coronograph with a FOV of $24\times$\ang{;;24} (as seen in Figure\,\ref{fig:miri_images}). For our eight pointings, this adds up to a total area of just under 19.4\,square arcminutes. The effective area we use is reduced to 18.4\,square arcminutes after masking out noisy edges and bright stars. This is discussed in Section\,\ref{subsec:masking}.


\begin{figure}
    \centering
    \includegraphics[width=\linewidth]{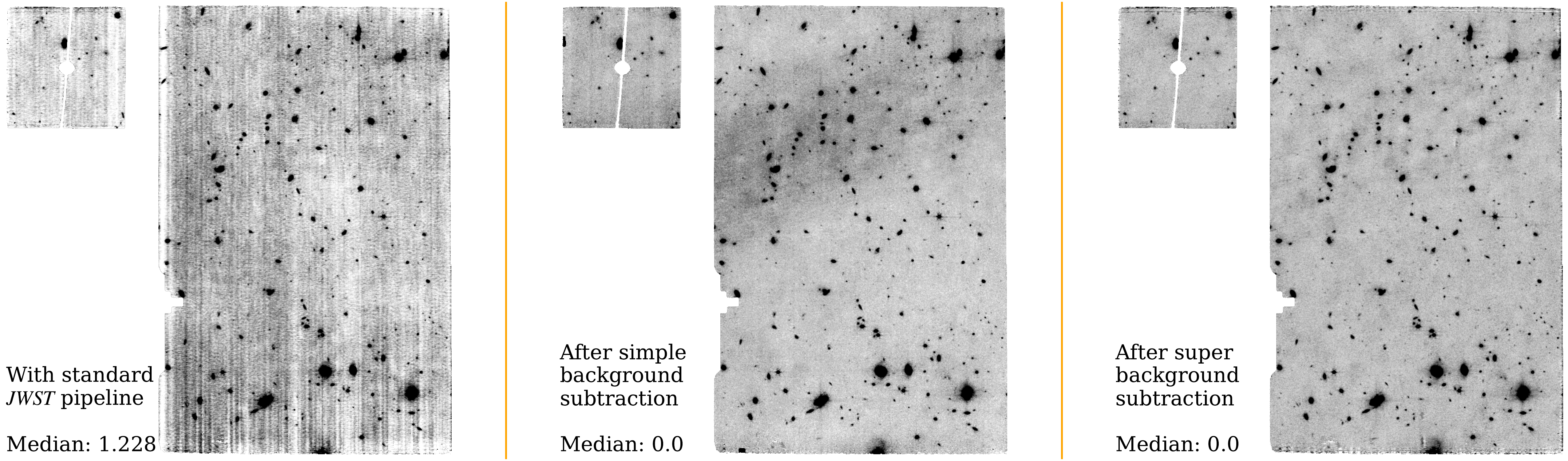}
    \caption{Example images produced with the standard pipeline, and following two different background removal methods, as described in Section\,\ref{subsec:data_reduction}. The image median (in counts/second/pixel) are given in the bottom-left of each panel. \textit{Left}: the FLS6 image from the standard pipeline. \textit{Middle}: the FLS6 image where we have subtracted a median background constructed from all six FLS fields, with the sources masked out. \textit{Right}: the FLS6 image with super background subtraction \citep{Perez-Gonzalez2024,Alberts2024}. }
    \vspace{12pt}
    \label{fig:bkg-sub}
\end{figure}


\subsection{Data reduction}
\label{subsec:data_reduction}

We first obtained our MIRI images from the archive already reduced using the {\sl JWST} operations pipeline build 10.1 (science calibration pipeline version 1.12.5; Calibration References Data System (CRDS) version 11.17.6, context jwst\_1193.pmap)\footnote{As of the time of writing, this was the latest pipeline build that affected MIRI imaging. See \url{https://jwst-docs.stsci.edu/jwst-science-calibration-pipeline-overview/jwst-operations-pipeline-build-information}}. The left panel of Figure\,\ref{fig:bkg-sub} shows an example of one of our images. Within these pipeline-reduced images we noted both the presence of an overall background level (the median pixel value for the FLS images was 1.23, for the COSMOS images it was 1.52). This difference in overall background level is unsurprising given that FLS is a northern field, whereas COSMOS is equatorial and as such sees a higher level of Zodiacal light, see example discussion in \citet{Sanders2007}. We also noted some clear striping especially toward the bottom quarter of each image. 

We first attempted a simple background subtraction motivated by the fact that the striping patterns were fairly consistent among the images. We thus constructed a median background image where we used the segmentation map provided with pipeline-processed images to mask out bright sources in the fields. The segmentation map is a copy of the original image\footnote{This map is produced for each image during the source extraction step of the pipeline processing, see Section\,\ref{subsec:sextractor} for details regarding source extraction. where pixels are replaced either with the value 0 if they are not part of an identified source or with a positive value if they are.} To avoid any remaining source signal, we added a 5 pixel ``buffer" around each source (positive-value pixel) within the segmentation maps. Due to the difference in overall pixel level between the images, we rescaled each source-subtracted image to a median of one. The middle panel of Figure\,\ref{fig:bkg-sub} shows an example of our image after this basic median background subtraction when we only use the FLS images for constructing the background. It is clear that this method works well in removing the vast majority of the striping though some residual structure remains due to second order differences between images. In addition, since we only have two COSMOS images we could not construct a COSMOS only median using this method. We attempted a combined median background in FLS and COSMOS; however, this led to the COSMOS fields (which have less weight in this median with 2/8 images only) to have higher noise levels. Specifically, using a single median we found 5\,$\sigma$ levels of $\approx \SI{0.095}{\micro Jy}$ in FLS and $\approx \SI{0.135}{\micro Jy}$ in COSMOS. When using only the FLS images for a median (as shown in the middle panel of Figure\,\ref{fig:bkg-sub}), we find 5\,$\sigma$ levels of $\approx \SI{0.085}{\micro Jy}$. While we accounted for the differences in overall background level, these results clearly indicate that we have some time variable aspects to the background which make it sub-optimal to construct such median background from observations about 6 months apart. 

To address these issues, we then processed our images through the super-background subtraction procedure developed by the SMILES (PID 1207; PI G. Rieke; \citet{Lyu2024}) team \citep{Perez-Gonzalez2024,Alberts2024}. This background subtraction starts with the stage 2 images and homogenizes the background across all images in the program taking into account the time varying behavior of the background which is particularly prominent at shorter wavelengths as here. The full details are described in \citet{Alberts2024}. The right panel of Figure\,\ref{fig:bkg-sub} shows the result of this super background subtraction. It is clear that the residual structure in our previous simple background-subtracted image is now largely gone. In addition, this method allows for COSMOS only median since it no longer uses the fully processed images (of which there are only two), but the stage 2 images of which we have 2$\times$6 dithers which is sufficient for a median background. As we show in  Section\,\ref{subsec:imagedepth} below, we also no longer find a difference in the obtained depth in FLS and COSMOS. The depth we find in FLS relative to the simple background subtraction using the FLS images alone however are comparable. This suggests that depending on the specific science applications and data available, a simple procedure as described above may be sufficient, but it should only be applied to images obtained close in time to avoid time-varying background effects.   

\subsection{Noise properties}
\label{subsec:noise}

\begin{figure}
    \centering
    \includegraphics[width=\linewidth]{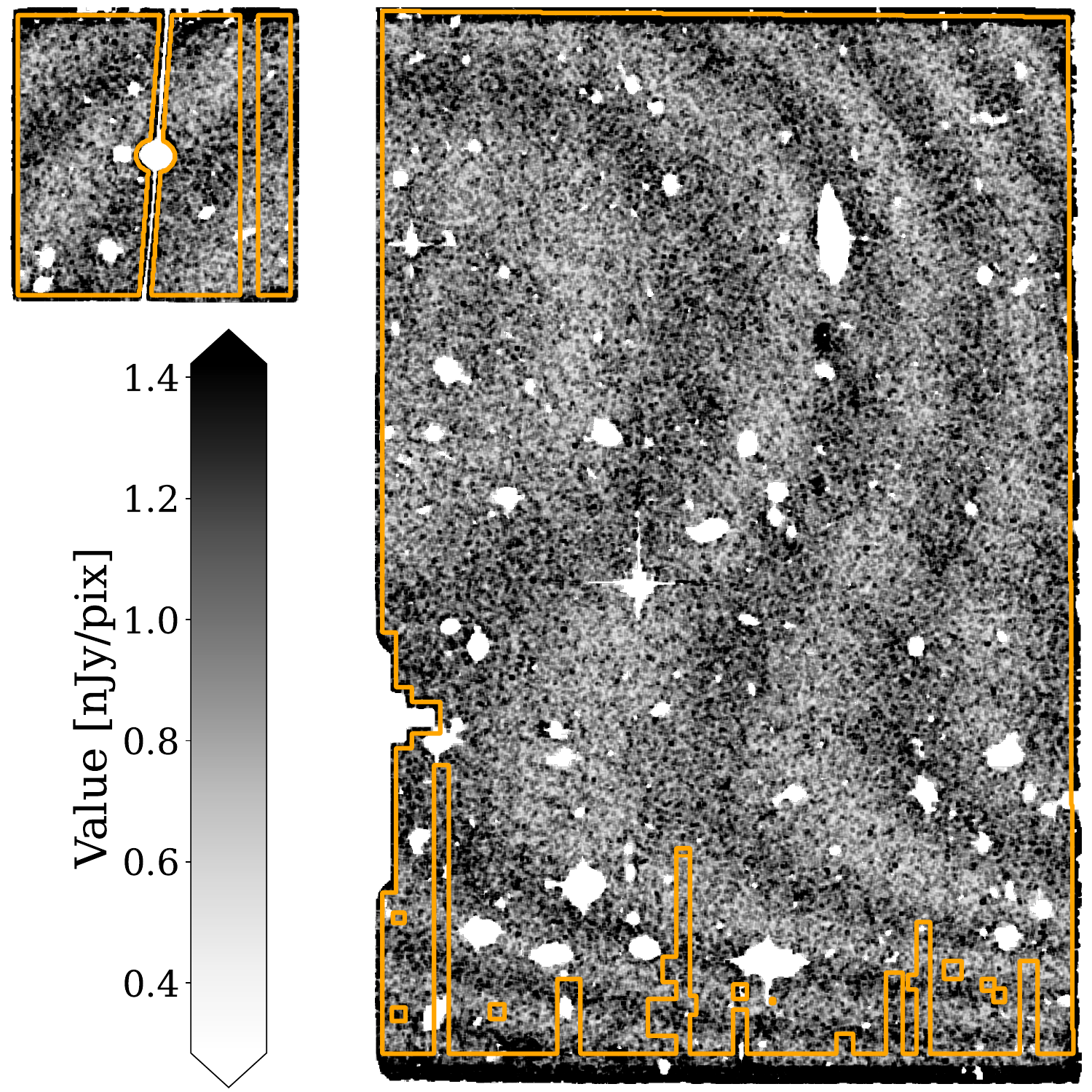}
    \caption{Noise map for the FLS1 field. Produced by computing the standard deviation of the FLS1 image inside a $3\times3$ pixel square centered on each pixel. The science image used in producing the noise map was first multiplied by the field's segmentation map, removing the sources. We then applied a mask, shown here in orange, that removes edges and residual stripes as additional sources of noise. After both steps, we produced the noise map. A thin orange line on the colorbar indicates the median value of the pixels in the noise map.}
    \label{fig:fls1-noise-map}
\end{figure}

To examine the noise characteristics of our images post background subtraction, we created a noise map. In this map, each pixel's value represents the standard deviation in the image pixel values in a 3$\times$3 pixel grid centered on that pixel. We performed this calculation on each image with two masks applied. First, we masked the pixels belonging to sources by applying the segmentation map. We then created a mask that removes the image edges and the residual strips, as we consider them to be sources of additional noise. This second mask is the same for each image. We perform our noise calculation on the science images after the segmentation map and noise/stripe mask have been applied. The resulting noise map for the FLS1 field is shown in Figure\,\ref{fig:fls1-noise-map}. We compared our produced noise map to the one provided with the standard pipeline products (which is derived from the per pixel coverage). The typical pixel values of these calculated noise maps are 60-70\% of the MAST-provided error maps. 

Figure\,\ref{fig:fls1-noise-map} shows a wavy pattern across the whole image. We found this pattern in all our images, even when varying the number of pixels used in calculating the standard deviation, and also when running the same procedure before or after background subtraction. We also found this pattern with both the simple and super background subtracted images. We did not find this pattern when running the same procedure on one of the publicly-available CEERS images in the same filter \citep{finkelstein2022long}. The CEERS images have lower exposure times (0.82 vs. 1.85\,hrs), and slightly smaller pixel scales (0.09$^{\prime\prime}$ vs 0.11$^{\prime\prime}$). At present, we consider this an unexplained instrumental effect that is consistent with the description of ``tree rings'' that have previously been observed in MIRI imaging\footnote{See \url{https://jwst-docs.stsci.edu/jwst-calibration-pipeline-caveats/jwst-imaging-pipeline-caveats} for details on the ``tree ring'' patterns.}. This pattern is at a very low level, with the amplitude of the peaks and troughs translating to $\approx$\,0.5\,nJy/pixel, as can also be seen in the scale color bar. 

\subsection{Image depth}
\label{subsec:imagedepth}
We estimate the image depth by placing random apertures in source-free parts of the background-subtracted images. The standard deviation of the Gaussian distribution of the resultant empty aperture fluxes is the $1\,\sigma$ depth of the image. The above method is a simplification of the method used in \citet{Annunziatella2023}, since we have fairly uniform coverage unlike that in \citet{Annunziatella2023} so noise-weighting is not critical. In  Section\,\ref{subsec:sextractor} we describe our \textsc{SExtractor} photometry. For our random apertures we use a diameter equal to the median of the \textsc{SExtractor} Kron radii, where a Kron radius is a ``reduced pseudo-radius'' that defines a circular aperture containing $\ge 90\%$ of the flux of the source\footnote{Refer to \url{https://sextractor.readthedocs.io/en/latest/Photom.html##automatic-aperture-flux-flux-auto} and section 6 of \citet{Bertin:1996} for details.}. We choose Kron apertures as they are meant to capture most of the light of our sources, which are typically spatially extended.  The median Kron radius in our catalog is $\sim \ang{;;0.385}$, so we adopt a diameter of $\ang{;;0.77}$ for our apertures. For comparison, the FWHM of the MIRI PSF at \SI{5.6}{\micro \meter} is \ang{;;0.207}.

We placed 500 empty apertures per field for a total of 4000 apertures. For each image, we then fitted a Gaussian to the resulting histogram of aperture flux. We found that the standard deviations of these fitted Gaussians were consistent between all the images, translating to a $5\,\sigma$ depth of \SI{0.1}{\micro Jy}.

As a cross-check, the median flux uncertainties for our sources based on the SExtractor photometry (Section\,\ref{subsec:sextractor}) translate to a 5\,$\sigma$ uncertainty of 0.081$\mu$Jy. This is reasonably consistent with our image depth analysis above. Note that the random apertures method is sensitive to the adopted aperture sizes and slightly smaller apertures would bring the two values into even closer agreement.  As an additional cross-check, the image depths quoted by CEERS \citep{Yang2023}, rescaled to our exposure time (since depth scales as $\sqrt{t}$) translate to $\approx \SI{0.08}{\micro Jy}$ for our images. To be conservative, we adopt $\approx \SI{0.1}{\micro Jy}$ as our nominal 5\,$\sigma$ image depth. 

\newpage

\subsection{Photometric zero-point} \label{subsec:psf}

Pre-launch estimates of the MIRI imager's performance already predicted that its point spread function (PSF) has a `cruciform' artifact that draws luminosity from the core into the wings of the Gaussian. In-flight measurements have shown that this artifact is prominent at shorter wavelengths like ours \citep{rigby:2023}, but is negligible at $> \SI{10}{\micro \meter}$ \citep{Gaspar2021}. The extent of this effect was only noted once in-flight assessments had been made. Since our data were obtained very early after the start of science operations, the FITS header keyword PHOTUJA2 provided for converting from the native units to physical photometry units was incorrect. The fiducial keyword translated to a photometric zero-point of 26.121 mag which resulted in us seeing a significant systematic offset in the photometry measured by MIRI F560W as compared to the \textit{Spitzer} IRAC Ch3 data available for our fields.

This issue was also noted by the CEERS team and led them to calculate a photometric zero-point for the \SI{5.6}{\micro \meter} images of 25.701 in AB magnitudes \citep{papovich:2023}.  However, the CEERS images have a pixel scale of 0.09$\arcsec$ \citep{papovich:2023} whereas ours is 0.11$\arcsec$. We re-scale the CEERS calculated zero-point as in $zp = 25.701 - 5 \log_{10}(0.11/0.09) = 25.265$. This is the value we adopt in this paper which leads to much better agreement between the MIRI\SI{5.6}{\micro \meter} and IRAC channel 3 photometry for cross-matched sources (see Section\,\ref{sec:verification}). 

\section{Data analysis}
\label{sec:analysis}

\subsection{SExtractor photometry} \label{subsec:sextractor}

We used \textsc{Source Extractor} \citep{Bertin:1996} version 2.28.0 for source detection and photometry. Table \ref{table:se_params} lists our \textsc{SExtractor} parameter settings. The SEEING$\_$FWHM keyword adopted here corresponds to the FWHM of the in-flight measured MIRI \SI{5.6}{\micro \meter} point spread function (PSF) (see section \ref{subsec:completeness} for details).  As our fiducial flux measurements, we adopt values produced with Kron aperture photometry, labeled by the \texttt{FLUX\_AUTO} keyword by \textsc{SExtractor}. The photometric uncertainties are natively provided by SE and take into account the variance of the pixel values inside the aperture, the background value, and the effective detector gain.

\begin{table}[h!]
\vspace{48pt}
\centering
\begin{tabular}{>{\raggedright}p{4.cm} c} 
 \hline\hline
 \textbf{Extraction} &  \\
    DETECT\_MINAREA & 9.0 \\
    DETECT\_THRESH & 1.5 \\
    ANALYSIS\_THRESH & 0.4 \\
    FILTER\_NAME & gauss\_3.0\_5x5.conv \\
 \hline
 \textbf{Photometry} & \\
    PHOT\_AUTOPARAMS & 2.5, 3.5 \\
    PHOT\_FLUXFRAC & 0.5 \\
    MAG\_ZEROPOINT & 25.265 \\
    GAIN & 36630.528 \\
    PIXEL\_SCALE & 0.11\\
 \hline
 \textbf{Background} & \\
    BACK\_TYPE & AUTO  \\ 
    BACK\_SIZE & 16  \\ 
    BACK\_FILTERSIZE & 3  \\ 
    BACKPHOTO\_TYPE &  LOCAL \\
 \hline
 \textbf{Deblending} & \\
    DEBLEND\_NTHRESH & 32 \\
    DEBLEND\_MINCONT & 0.003 \\
 \hline
 \textbf{Star/Galaxy Separation} & \\
    SEEING\_FWHM & 0.207 \\
 \hline\hline
\end{tabular}
\caption{\textsc{SExtractor} parameter settings for the photometry.}
\label{table:se_params}
\end{table}

\newpage
\subsection{Masking} \label{subsec:masking}

We further clean the raw \textsc{SExtractor} catalogs by masking out bright stars as well as any particularly noisy parts of the images.
For each image, we created masks that include two components. One excludes obviously diffracted sources. The other avoids image edges\footnote{The edges are masked since, besides the issue of sources falling partly outside the image area, we have higher backgrounds there (see Figure\,\ref{fig:miri_images})  largely due to the lower exposure times on the edges since the primary MRS observations involved a small dither, see Young et al. (2023, in prep).} where we assume an edge width of 5 pixels. The masks also include the blocked regions of the Lyot coronograph) and 
the residual striping artifacts -- most of these were removed in the background subtraction step but some residual striping remained near the bottom of the images, which we define manually (see right-hand panel in Figure\,\ref{fig:bkg-sub}). Figure\,\ref{fig:fls1-noise-map} shows as an example the mask for the FLS1 field.

We also apply a quality cut on our catalogs, only keeping sources with a signal-to-noise ratio greater than 5. The combination of the masking and the quality cut reduces the raw \textsc{SExtractor} output of 4,089 sources to a total of 2,854 ``reliable" sources (``reliable" meaning falling outside our defined masks and having SNR$\ge5$) among all our eight fields. 
The masking improves the reliability of our source population, since within the masked areas we find a significantly larger fraction of fake sources, relative to the rest of the images as discussed in Section\,\ref{subsec:fakesources}, due to ``sources" being picked up by \textsc{SExtractor} on top of strong diffraction spikes or stripes.  Per field, after masking we detect the most sources in field FLS3 (409 sources) and the least in field FLS5 (310 sources).

\begin{figure}
    \centering
    \includegraphics[width = \linewidth]{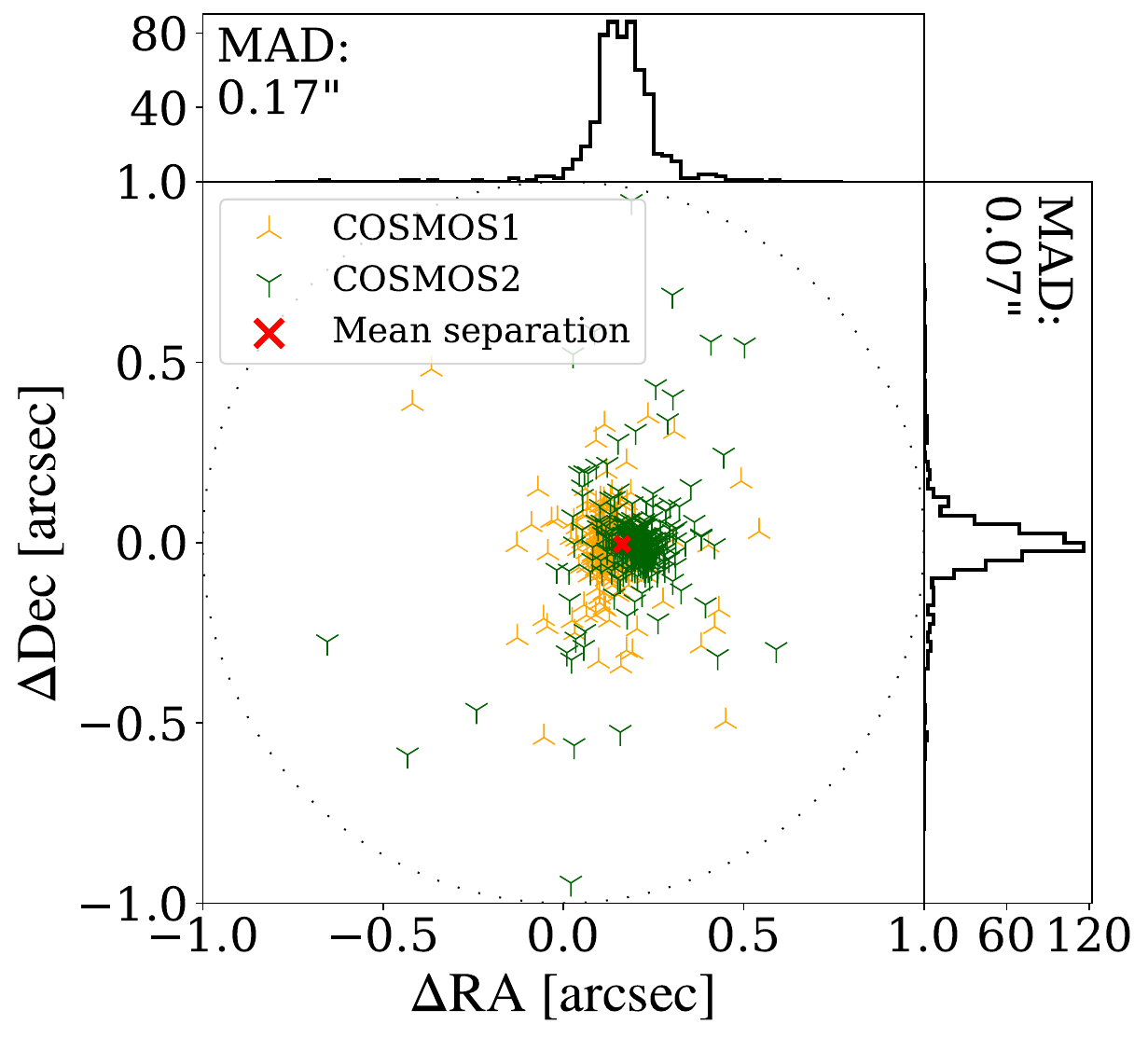}
    \caption{Comparison of the position centroids of the 582 matched sources between MIRI and the COSMOS2020. The histograms (top and right) represent the offsets in both fields combined, with the mean absolute deviation (MAD, calculated as the mean of $| \Delta \mathrm{RA} |$ and $| \Delta \mathrm{Dec} |$, respectively) labeled.}
    \label{fig:ra-dec-offsets}
\end{figure}

\subsection{Astrometric \& photometric verification \label{sec:verification}}

In order to perform astrometric and photometric verifications, we used the \textsc{SExtractor} detected sources in our two COSMOS fields and cross-matched them with the COSMOS2020 CLASSIC catalog \citep{Weaver2022}. The COSMOS2020 source detection is based on a weighted combined $izYJHK$ image \citep{Weaver2022}. We used a \ang{;;1} matching radius and found a total of 582 COSMOS2020 cross-matches, 175 of which also have IRAC channel 3 detections. 
We choose a \ang{;;1} radius as reasonable in a cross-match with the COSMOS2020 ground-based optical to near-IR data. As we see below, the vast majority of the matched sources have positional separations within \ang{;;0.5} so increasing the matching radius would not alter the results significantly while increasing the likelihood of spurious matches. 
Overall, our MIRI data increase the fraction of COSMOS2020 with 5.6/\SI{5.8}{\micro \meter} data by roughly 3 times. Note that our MIRI fields are within the deep part of the COSMOS2020 catalog and not near the edges and the COSMOS2020 positions rely on {\sl Gaia} astrometric solutions \citep{Weaver2022}. Figure \ref{fig:ra-dec-offsets} shows the comparison between the MIRI and COSMOS2020 source positions of the cross-matched sources. The offsets we find are small and well within the MIRI PSF at this wavelength. Therefore we do not explicitly apply astrometric corrections. 

\begin{figure}
    \centering
    \includegraphics[width = \linewidth]{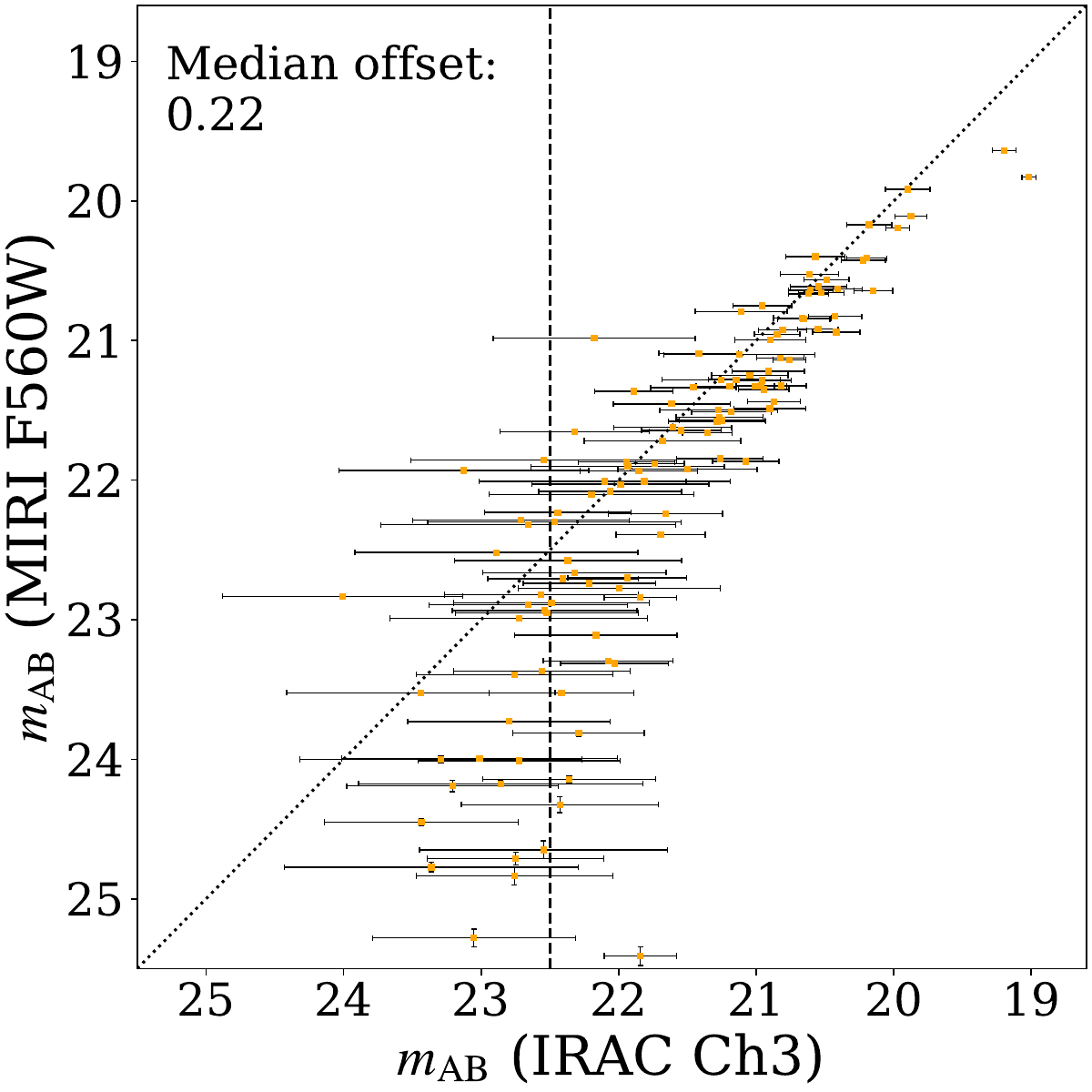}
    \caption{Photometry cross-check for the matched sources with available IRAC channel 3 measurements applying a $\mathrm{SNR} > 1$ cut on the IRAC ch3 data, leaving us with 119 sources. The dashed diagonal line is the 1:1 line. 
    The vertical dashed line represents the $3 \sigma$ IRAC channel 3 limit \citep{Weaver2022}. The quoted median offset is computed for the SNR$>3$ sources. }
    \label{fig:photometry-crosscheck}
\end{figure}

Figure \ref{fig:photometry-crosscheck} shows the comparison between the MIRI \SI{5.6}{\micro \meter} and the available IRAC channel 3 fluxes. As described in section \ref{subsec:sextractor}, we use Kron aperture photometry for the \SI{5.6}{\micro \meter} fluxes. For the IRAC fluxes, we used the aperture corrected \ang{;;2} aperture photometry from \citet{Weaver2022_catalog}. We overlay the 1:1 line. 
Overall, we find good agreement, although we have a median offset of 0.22 magnitudes. We compare this to the results presented in Figure 8 of \citet{Yang2023}, who have significantly deeper IRAC data and therefore are able to compute the median offset in magnitude bins down to $\approx$25th magnitude. Their median offset at their IRAC 5\,$\sigma$ bin of $\approx$22.5 magnitude is roughly 0.1$\pm$0.1, and gets much larger at fainter magnitudes and lower SNR. Their offset is near zero at magnitudes brighter than $\approx$21. Most of our sources with IRAC SNR$>$3 are in the 21-22.5 magnitude range. Here we are consistent with \citet{Yang2023} in that we find a positive offset (i.e. generally the MIRI magnitudes are fainter) and our median offset is within 1\,$\sigma$ of theirs for the $\approx$22.5 magnitude bin. 

\begin{figure}
    \centering
    \includegraphics[width = \linewidth]{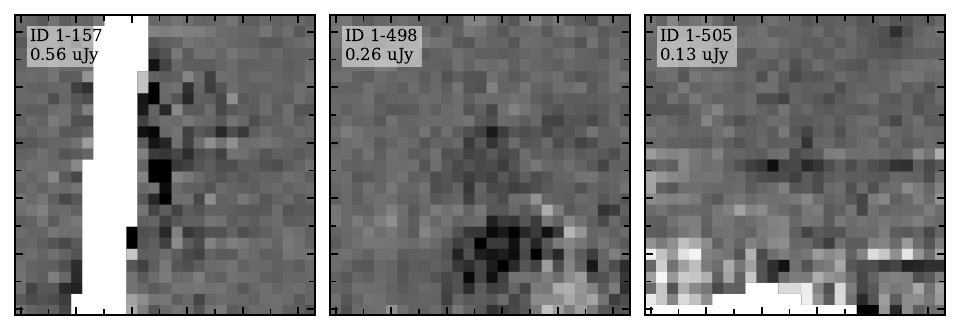}
    \caption{Representative examples of sources we flagged as ``fake", see Section\,\ref{subsec:fakesources}. The stripe in the left-hand panel is due to the mask around the Lyot coronograph. See text for further details re fake source flagging.}
    \label{fig:fake-sources}
\end{figure}

\subsection{Estimating fraction of fake sources \label{subsec:fakesources}}

\begin{figure}
    \centering
    \includegraphics[width = \linewidth]{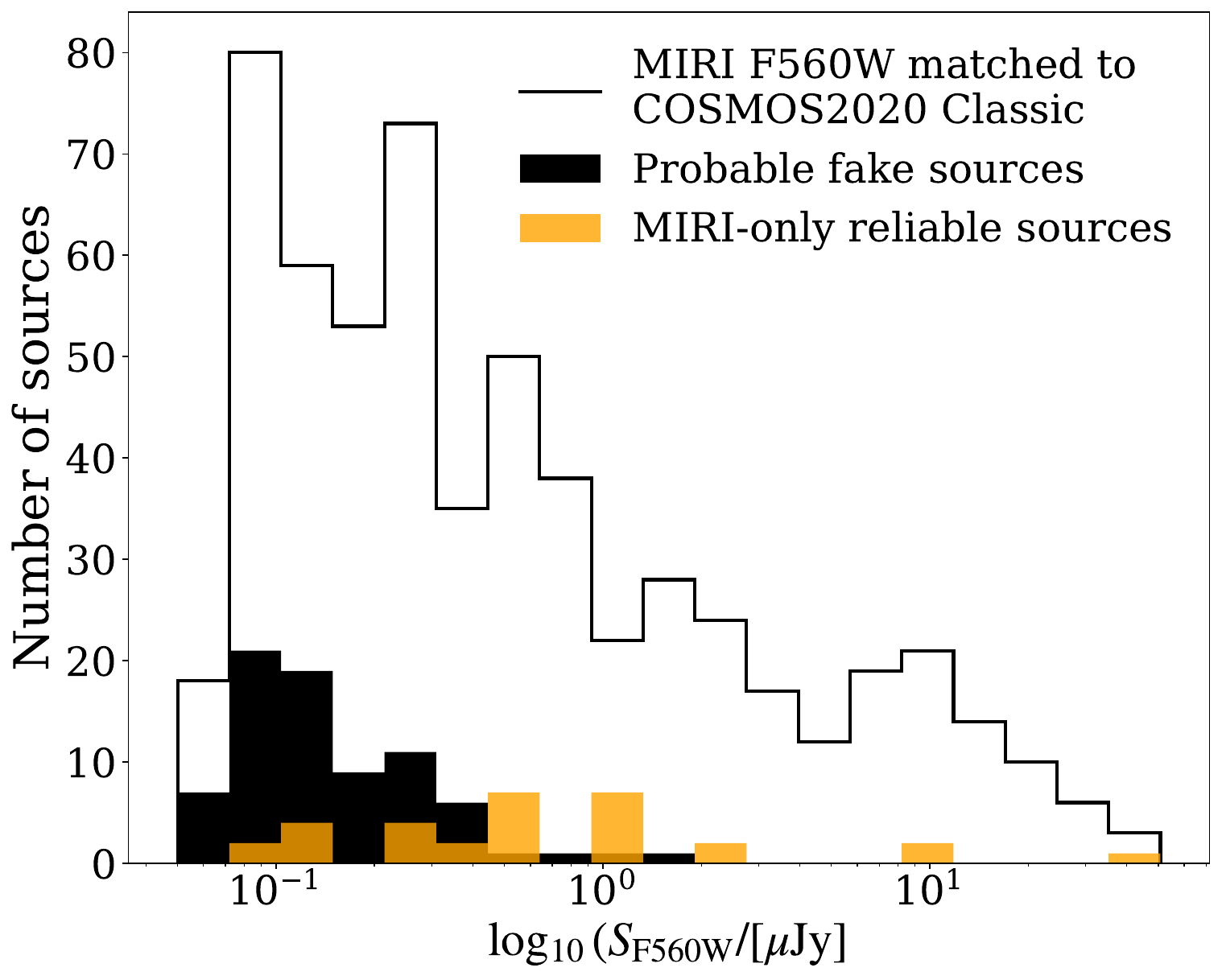}
    \caption{A comparison between the flux distributions of our MIRI sources that are matched to COSMOS2020 sources vs. those that are unmatched. }
    \label{fig:flux_distributions}
\end{figure}

The cross-match with COSMOS2020 also gives us an upper limit on the fraction of the \textsc{SExtractor} detected sources that might be fake. Such sources are expected due to some of the residual artifacts in the images, that may have been missed in the masks described in Section\,\ref{subsec:masking}. Overall we found COSMOS2020 counterparts for 582 of the 690 MIRI sources in our two COSMOS fields (84\,\%). The 108 unmatched sources ($16\,\%$), represent an upper limit on the fake source fraction since it is expected that we have some red MIRI sources that are not detected in COSMOS2020. 
Figure\,\ref{fig:flux_distributions} shows the \SI{5.6}{\micro \meter} flux distributions of the sources matched to COSMOS2020, compared to those of unmatched sources.
Unsurprisingly, the unmatched sources are generally fainter than the overall MIRI source population. 
Indeed, visual examination of the 108 unmatched sources showed only 31 are unambiguous MIRI detections. Removing these sources which are clean MIRI detections despite being unmatched in COSMOS2020 leaves us with 77 likely fake sources ($\sim10\%$ of the total). Some of these apparently fake sources are driven by single bright pixels that were missed in the standard pipeline reduction, others appear to be residuals of the striping pattern seen in Figure\,\ref{fig:bkg-sub} or other image artifacts. Figure\,\ref{fig:fake-sources} shows examples of sources we flagged as fake. In Figure\,\ref{fig:flux_distributions} we show separately the flux distributions of the unmatched sources judged to be real vs. those that are potentially fake. Note that Section\,\ref{subsec:veryredmiri} explores in more detail the nature of the 31 reliable MIRI sources that are unmatched in COSMOS2020.

\section{Results}
\label{sec:results}

\subsection{Source completeness} \label{subsec:completeness}

In order to estimate the source detection completeness at different flux density levels, we take the usual Monte Carlo approach by injecting fake sources at different flux levels and noting the fraction of them that are recovered with the same \textsc{SExtractor} procedure as applied to real sources \citep[e.g.][]{Takagi:2012}. We use the code described in \cite{Shipley:2018} for this step. 

\begin{figure}[h]
    \centering
    \includegraphics[width = 0.75\linewidth]{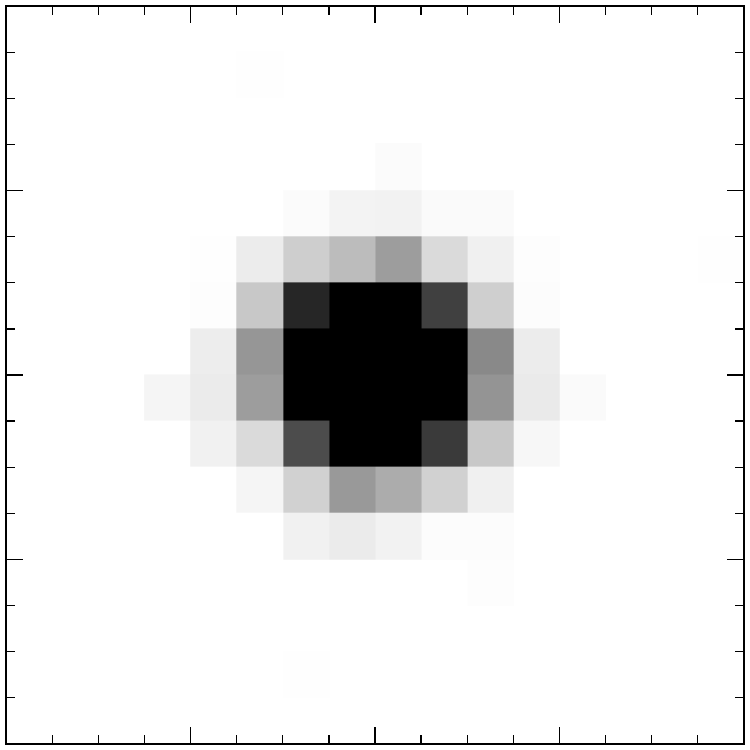}
    \\
    \caption{The $16 \times 16$ pixel source model used in the Monte Carlo simulations to determine source completeness. The FWHM of the best-fit Gaussian is $\ang{;;0.341}$ in the x-direction and $\ang{;;0.330}$ in the y-direction.}
    \label{fig:psf}
\end{figure}

For the injected fake sources, we need a model of their on-the-sky distribution. Simply adopting the MIRI PSF, is not appropriate since the vast majority of our sources are spatially resolved. To create a more realistic source model, we stack 500 isolated and visually compact sources to construct a source model. Figure\,\ref{fig:psf} shows this model which measures 16 pixels $\times$ 16 pixels ($\approx \ang{;;1.8} \times \ang{;;1.8}$). This source model has a mean (along x- and y-axis) full width at half-maximum (FWHM) of \ang{;;0.357} which, as expected, is larger than the MIRI \SI{5.6}{\micro \meter}m PSF of \ang{;;0.207}, as measured during \textit{JWST} commissioning\footnote{For details, refer to \textit{JWST} user documentation at: \url{https://jwst-docs.stsci.edu/jwst-mid-infrared-instrument/miri-performance/miri-point-spread-functions}.}.

\begin{figure}[h]
    \centering
    \includegraphics[width = \linewidth]{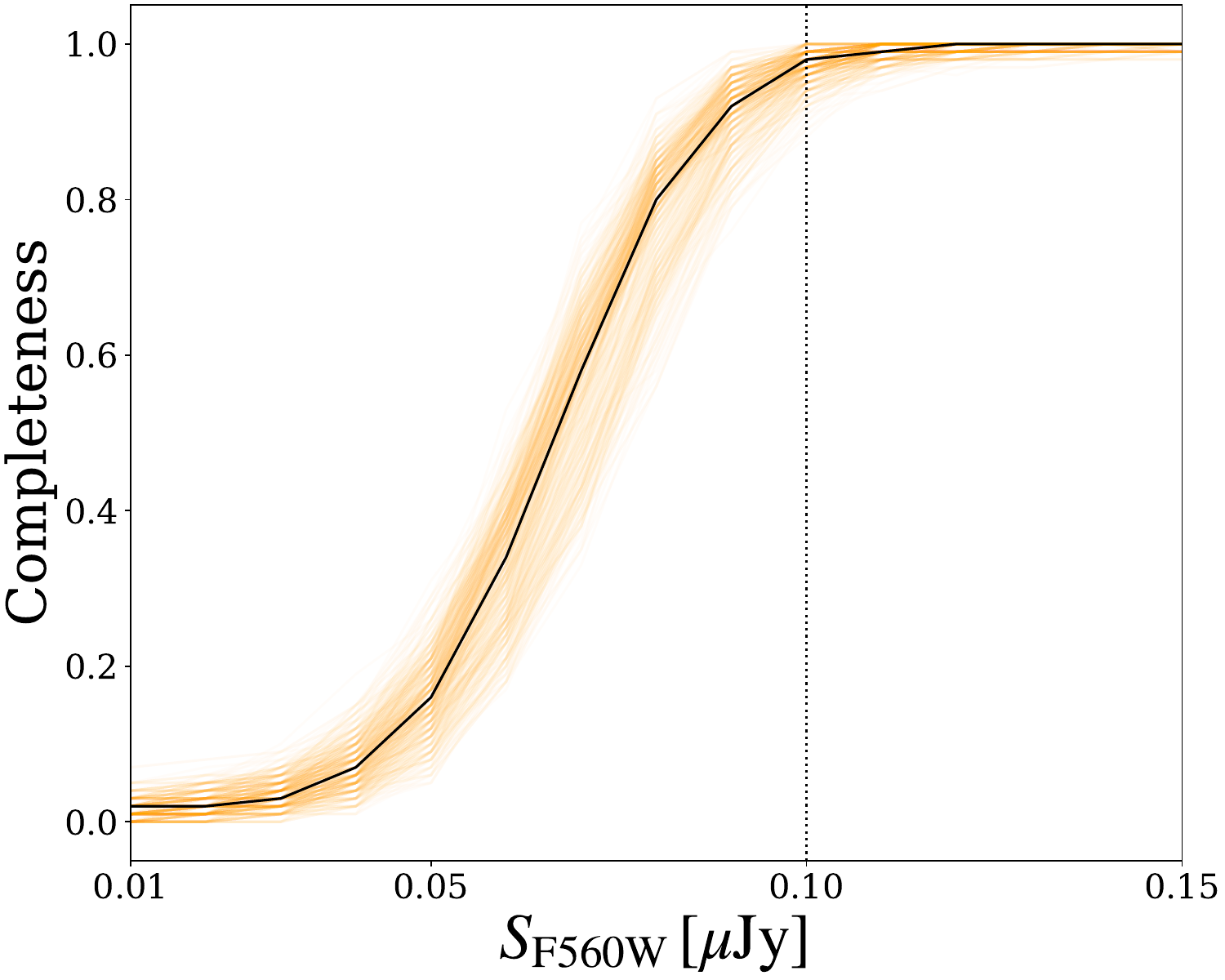}
    \caption{Source completeness for our images. The $x$-axis is trimmed to show detail. We overlay, as thin orange lines, an arbitrary sample of the compututed completeness curves in order to show their spread. 
    The vertical dotted line shows the $5 \sigma$ depths of the images as determined by the empty apertures method, detailed in section \ref{subsec:imagedepth}.}
    \label{fig:completeness}
\end{figure}

Completeness estimation starts with a segmentation map produced by \textsc{SExtractor} where all pixels with zero value correspond to no source present. We additionally set all pixel values to zero if they were masked, as described in Section\,\ref{subsec:masking}, as well as setting to zero all sources consisting of 25 or fewer contiguous pixels. This second step was done in order to provide more sites suitable for injection, as our procedure avoids injecting in sites flagged by the segmentation map as having sources, though we deem this unnecessary for smaller sources. We then proceeded to inject fake sources across a range of flux densities ranging between $\SI{0.01}{\micro Jy}$ and $\SI{0.25}{\micro Jy}$ in 25 total increments. Sources are injected only on valid pixels not removed in the above steps. At each flux level, we injected 500 sources in the image. The sources are injected by using the source model scaled to the desired flux density. We then processed the newly-created images with the same \textsc{SExtractor} procedure used for the science images themselves. The ratio of recovered to injected sources at each flux density level constitutes the completeness at that level. This procedure was done 10 times per flux level per image, totaling 80 runs. Figure\,\ref{fig:completeness} shows the measured completeness curves for each run as well as the adopted completeness curve, which is the median between the 80 runs. As expected we find that the completeness is essentially 100\% above our estimated 5\,$\sigma$ image depth and drops rapidly below that. The 50\% completeness is at $\approx \SI{0.07}{\micro Jy}$.  

\begin{figure*}
    \centering
     \includegraphics[width=0.45\linewidth]{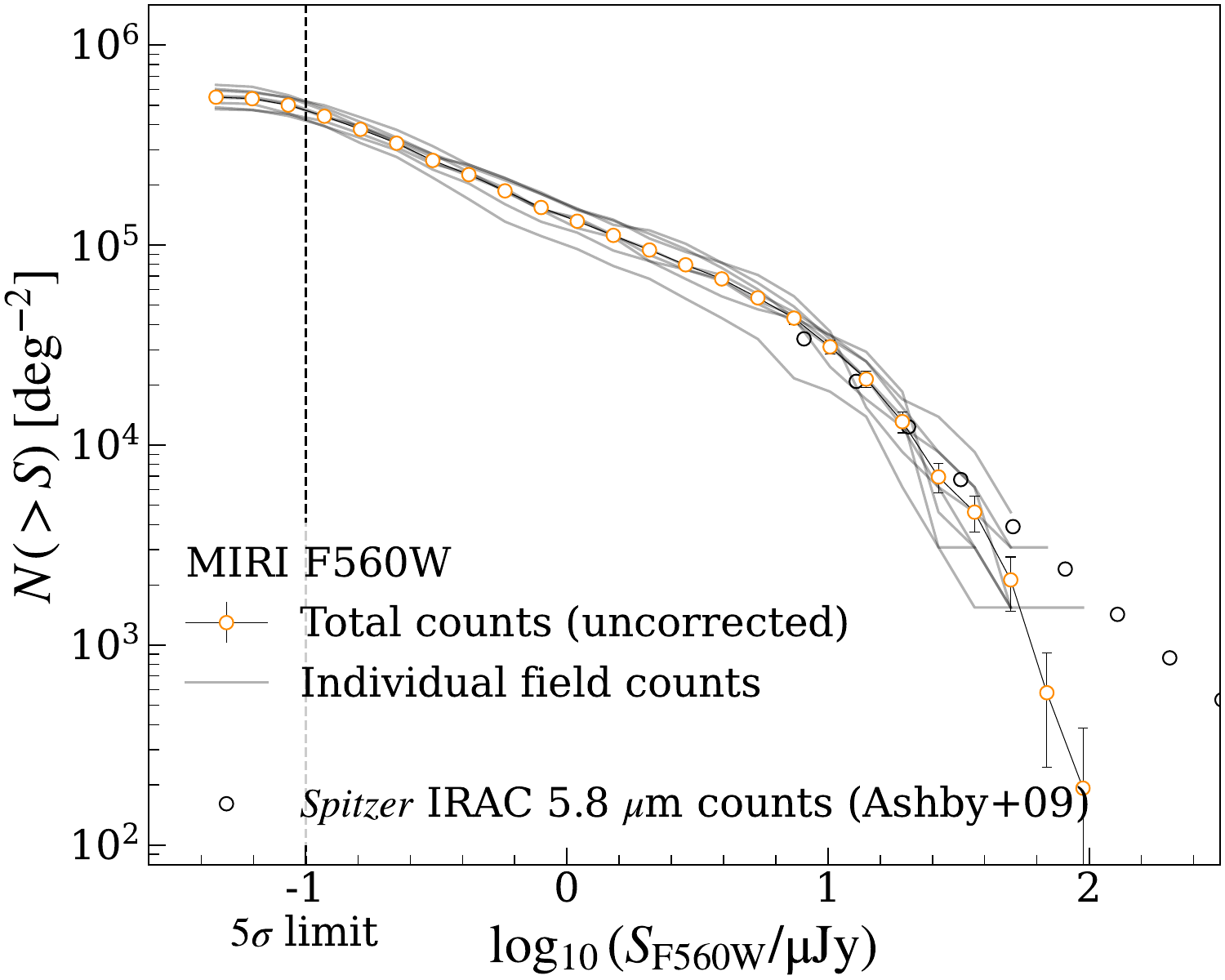}
    \includegraphics[width=0.45\linewidth]{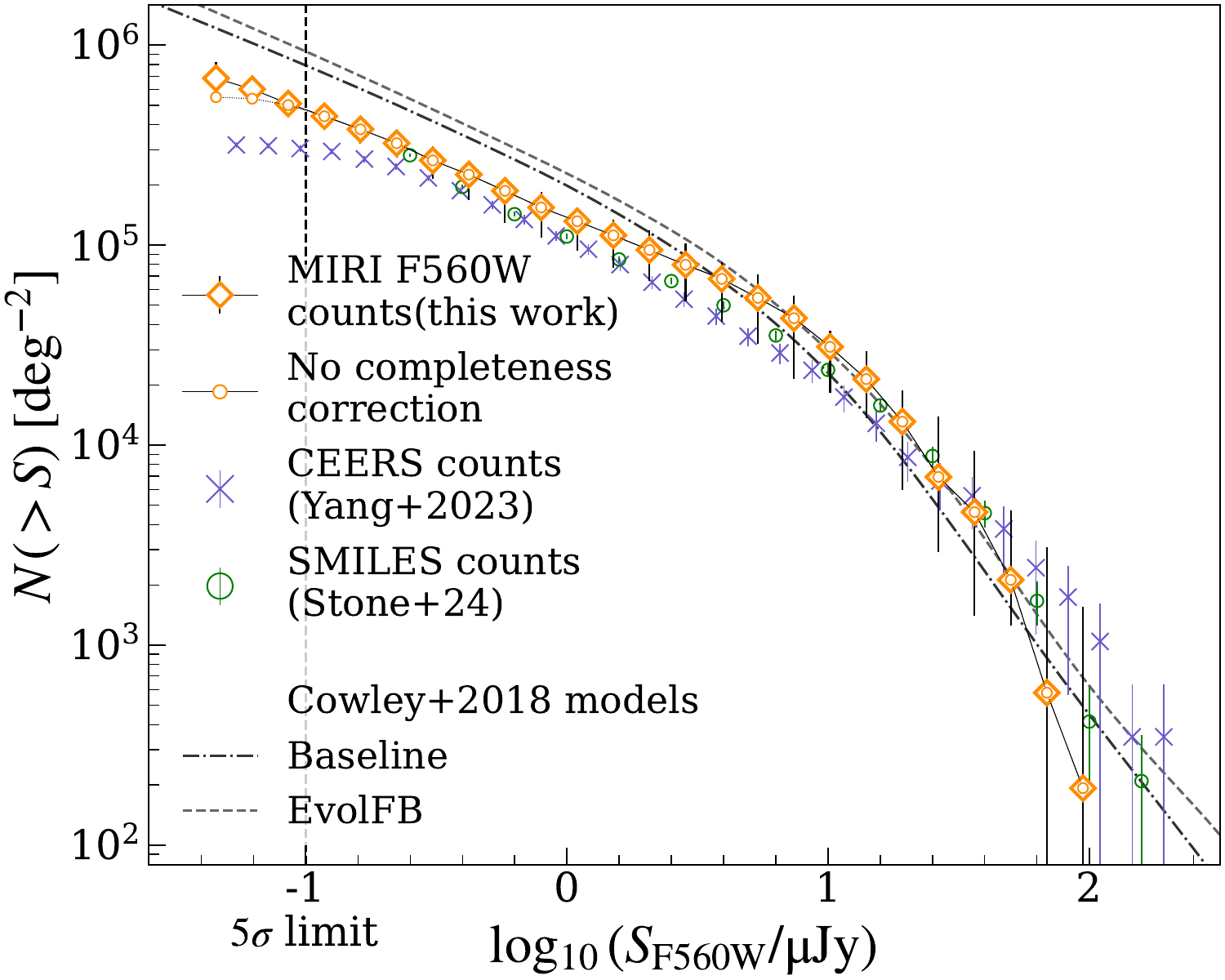}
    \caption{\textit{Left} The combined integral number counts with Poisson errors along with the counts for our individual eight fields. The field-to-field variation is roughly a factor of two -- well about the Poisson errors on the combined counts. The vertical dashed line is our  $5\,\sigma$ limit. Lastly, we overlay the IRAC channel3 counts from \citet{Ashby2009}. These do not have the stars removed which explains the difference between the counts at the bright end. \textit{Right} Our integral number counts for our eight fields (orange symbols), with the Poisson and field-to-field variation uncertainties added in quadrature. We overlay for comparison both the published CEERS counts \citep[][blue crosses, include only Poisson error]{Yang2023} and the SMILES counts \citep[][green circles, include only Poisson error]{Stone2024}, as well as model predictions MIRI F560W counts from \citet[][black curves]{cowley:2018} where the dot-dash curve is their fiducial model and the upper (dashed) one is the same but with evolving supernovae feedback (``EvolFB''). We also overlay the $5\,\sigma$ limit as a vertical dashed line.}
    \label{fig:number-counts}
\end{figure*}

\subsection{Number counts \label{subsec:number_counts}}

We calculate the MIRI \SI{5.6}{\micro \meter} integral number counts by first counting the number of objects in each flux bin $i$ as $dN/dS_i$. The source completeness for each flux bin is $C_i$ and is as shown in Figure\,\ref{fig:completeness}. The integral counts are given by:

\begin{equation}
N(>S)=\sum_{S_i=S}^{\infty} \frac{1}{A_{\mathrm{eff}, i}} \frac{1}{C_i}\frac{dN}{dS_i}
\label{eqn:counts}
\end{equation}

We calculate the effective area, $A_{\mathrm{eff}}$, as the inverse of the Boolean mask used for source detection (Section\,\ref{subsec:masking}) times the single pixel area. For all eight fields combined, this adds up to 18.4\,square arcminutes. Note that our coverage is fairly uniform within these masks therefore we assume the effective area does not change with flux level. The calculated integral counts are given in Table\,\ref{table:counts} along with their Poisson uncertainties. To calculate the raw counts, i.e. the ones not corrected for source completeness, we simply adopt $C_i\equiv 1$ in all flux bins in Equation\,\ref{eqn:counts}. When computing our number counts, we use less stringent quality cut on our catalog, counting all sources with an SNR $\ge 1$. In Figures\,\ref{fig:number-counts} we overlay the $5 \sigma$ limit of $\approx \SI{0.1}{\micro Jy}$ on our counts figures as a reference point. 

The left panel of Figure\,\ref{fig:number-counts} shows our combined MIRI \SI{5.6}{\micro \meter} integral number counts, with Poisson errors, along with the counts estimated for each of our eight fields individually. It is obvious that the field-to-field variation is significantly greater than the Poisson uncertainties on the combined counts. The shape of the counts is fairly consistent with nearly all fields showing a knee at around 6-\SI{8}{\micro Jy}. 

The right panel of Figure\,\ref{fig:number-counts} shows the combined integral MIRI \SI{5.6}{\micro \meter} number counts where the uncertainties on the points represent the total uncertainties. The latter are estimated by adding in quadrature the Poisson uncertainties with a field-to-field variation uncertainty which is taken to be half of the total spread between the individual field counts at each flux level. The combined counts and their Poisson and total uncertainties are given in Table\,\ref{table:counts}. 

For comparison, in Figure\,\ref{fig:number-counts} we overlay the CEERS $\SI{5.6}{\micro \meter}$ counts from \citep{Yang2023}. The data for the CEERS number counts come from four nearby fields (separated by $< \ang{;15;}$) with a total area of $\sim 9.5\,$square arcminutes. The median $5 \sigma$ depth for the CEERS images at $\SI{5.6}{\micro \meter}$ is \SI{0.138}{\micro Jy} \citep{Yang2023}. We note that our counts tend to be higher than the CEERS ones. This disagreement is only at $\approx$1\,$\sigma$ level for most flux bins considering our total uncertainties that account for field-to-field variation. The offset between our counts and the CEERS ones suggests the CEERS fields might be slightly underdense. Our widely separated fields allow us both an estimate of the average counts as well as a sense of the field-to-field variation. For example, in the left panel of Figure\,\ref{fig:number-counts}, some of our fields do show \SI{5.6}{\micro \meter} number counts consistent with those in the CEERS fields. Note that the uncertainties plotted on the CEERS counts are purely Poisson and do not include field-to-field uncertainties. The gap between our counts and theirs only becomes significant at the lowest flux bins, below $\approx \SI{0.3}{\micro Jy}$; the widest part of the gap is $\sim 2.3 \sigma$. We also overlay the just released MIRI \SI{5.6}{\micro \meter} counts from the SMILES team \citep{Stone2024}. They are in good agreement with our counts at the bright end, but closer to CEERS at the faint end, though again within our 1\,$\sigma$ uncertainties. We note that both CEERS and SMILES use combined F560W and F770W
images for source detection, unlike this analysis which is done entirely with MIRI \SI{5.6}{\micro \meter} images. To test the potential effect of differences in photometry, we ran our \textsc{SExtractor} setup on the public background-subtracted CEERS images and compared with their photometric catalog. We find excellent agreement, with the largest discrepancies per flux bin (e.g. for $<\SI{1}{\micro Jy}$ fluxes) being that we find roughly 10\% more sources. This is consistent with our estimated fake source fraction in Section\,\ref{subsec:fakesources}. Such fake sources are less likely in the CEERS photometric catalog due to their including the F770W data as well for source detection. This accounts for some of the offset we see at fainter fluxes, but $\approx$10\% likely fake source fraction is far from sufficient to explain the gap which is $\approx$2\,$\times$ at its maximum. Our Figure\,\ref{fig:number-counts}\,\textit{left} suggests this is primarily due to field-to-field variation instead.     

In the right panel of Figure\,\ref{fig:number-counts} we also overlay the model counts from \citet{cowley:2018}. Their predictions are based on the GALFORM semi-analytic model for galaxy formation within a $\Lambda$CDM framework \citep{Lacey2016}. The predictions based on the core galaxy formation model from \citet{Lacey2016} are referred to as the "Baseline model". The model that includes supernovae feedback whose strength evolves with redshift is called the ``EvolFB model''\footnote{We obtained the \citet{cowley:2018} model data from \url{http://icc.dur.ac.uk/data/}.}. The GALFORM model assumes the \citet{Maraston2005} stellar population models and the dust radiative transfer comes from the GRASIL prescription of \citet{Silva1998}. Our counts, in the brighter regime, are reasonably consistent with these model predictions with a slight preference for their evolving feedback model, although the difference between their models is well within our uncertainaties (including the field-to-field variation). In agreement with the CEERS and SMILES results, we find that the \citet{cowley:2018} models overestimate the counts for sources below $\approx \SI{3}{\micro Jy}$. 

\begin{deluxetable*}{ccccc}
\tabletypesize{\footnotesize}
\tablecolumns{5}
\tablewidth{0pt}
\tablecaption{The MIRI \SI{5.6}{\micro \meter} number counts \label{table:counts}}
\tablehead{
\colhead{log($S_{560W}$)} & \colhead{Raw $N(>S)$} & \colhead{Corrected $N(>S)$} & \colhead{Poisson error} & \colhead{Total error lower-upper\tablenotemark{$\dagger$}}  \\
\colhead{[\SI{}{\micro Jy}]} & \colhead{[10$^{-5}$deg$^-2$]} & \colhead{[10$^{-5}$deg$^-2$]} &  \colhead{[10$^{-5}$deg$^-2$]} &\colhead{[10$^{-5}$deg$^-2$]}}
\startdata
0.05 & 5.49 & 6.82 & 0.10 & 1.37-1.40 \\
0.07 & 5.40 & 6.01 & 0.10 & 0.96-1.05 \\
0.09 & 5.01 & 5.10 & 0.10 & 0.63-0.60 \\
0.12 & 4.41 & 4.41 & 0.09 & 0.54-0.55 \\
0.16 & 3.79 & 3.79 & 0.09 & 0.58-0.55 \\
0.22 & 3.23 & 3.23 & 0.08 & 0.49-0.53 \\
0.31 & 2.65 & 2.65 & 0.07 & 0.50-0.46 \\
0.42 & 2.25 & 2.25 & 0.07 & 0.58-0.28 \\
0.58 & 1.87 & 1.87 & 0.06 & 0.58-0.30 \\
0.80 & 1.54 & 1.54 & 0.05 & 0.45-0.30 \\
1.10 & 1.32 & 1.32 & 0.05 & 0.38-0.21 \\
1.51 & 1.12 & 1.12 & 0.05 & 0.35-0.23 \\
2.07 & 0.94 & 0.94 & 0.04 & 0.29-0.24 \\
2.85 & 0.80 & 0.80 & 0.04 & 0.28-0.22 \\
3.92 & 0.68 & 0.68 & 0.04 & 0.26-0.14 \\
5.39 & 0.54 & 0.54 & 0.03 & 0.22-0.17 \\
7.41 & 0.43 & 0.43 & 0.03 & 0.22-0.13 \\
10.18 & 0.31 & 0.31 & 0.02 & 0.13-0.06 \\
14.00 & 0.21 & 0.21 & 0.02 & 0.08-0.08 \\
19.25 & 0.13 & 0.13 & 0.02 & 0.07-0.08 \\
26.47 & 0.07 & 0.07 & 0.01 & 0.04-0.07 \\
36.39 & 0.05 & 0.05 & 0.009 & 0.03-0.05 \\
50.04 & 0.02 & 0.02 & 0.006 & 0.0086-0.026 \\
68.80 & 0.0058 & 0.0058 & 0.003 & 0.010-0.025 \\
94.60 & 0.0019 & 0.0019 & 0.002 & 0.014-0.014 
\enddata
\tablenotetext{\dagger}{These uncertainties include Poisson as well as field-to-field variation.}
\end{deluxetable*}

\begin{figure*}
    \centering
    \vspace{36pt}
    \includegraphics[width = \linewidth]{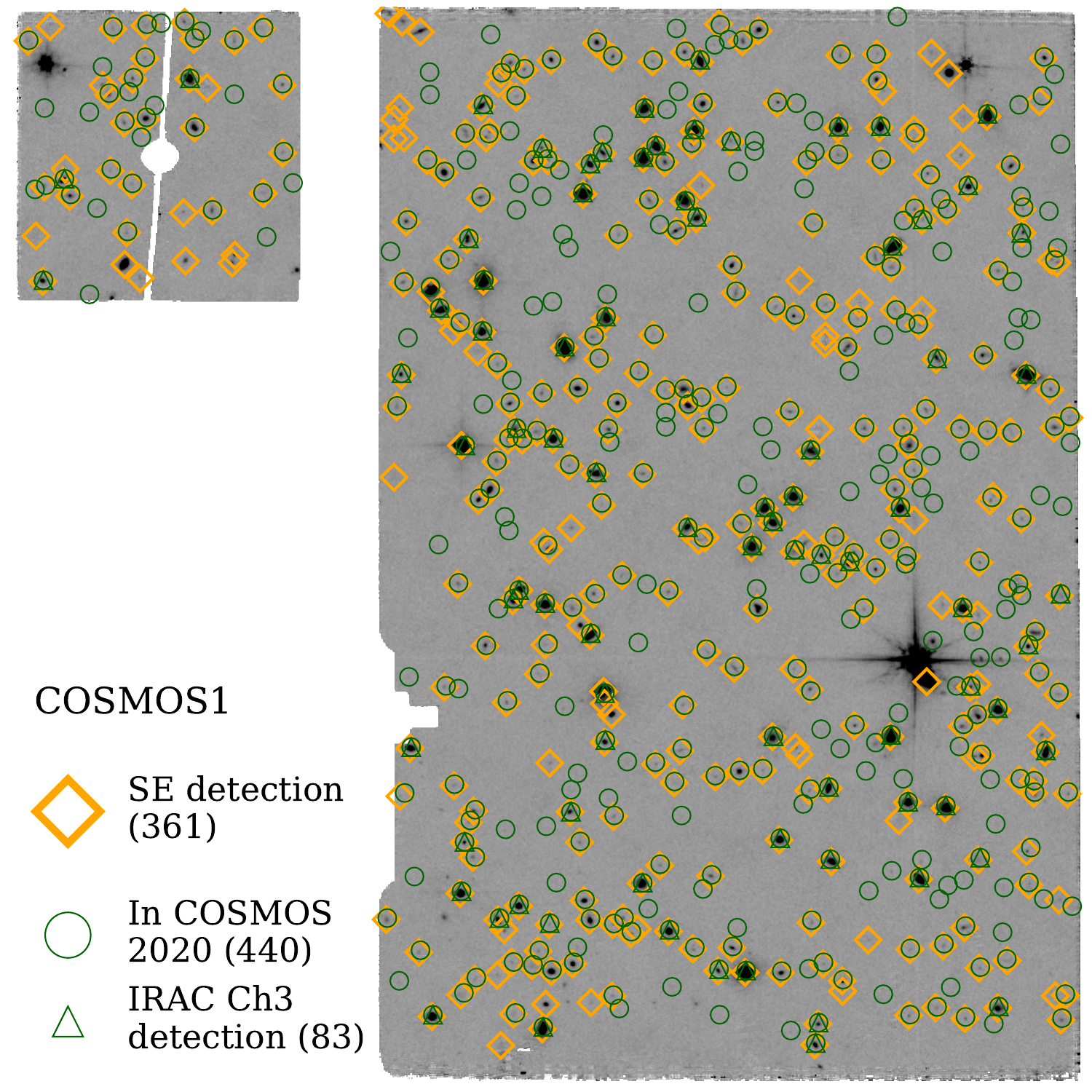}
    \caption{The COSMOS1 field with all COSMOS2020 Classic catalog \citep{Weaver2022} sources overlaid as the green circles. The COSMOS2020 sources which have IRAC channel 3 (\SI{5.8}{\micro \meter}) detections as shown as green triangles. The orange diamonds are all our MIRI-detected sources in this field. Note that we have $>$3$\times$ more MIRI \SI{5.6}{\micro \meter}m detections in the field relative to the prior \textit{Spitzer} IRAC \SI{5.8}{\micro \meter} detections.}
    \label{fig:crosscheck}
\end{figure*}
\clearpage


\begin{figure}
    \centering
    \includegraphics[width = \linewidth]{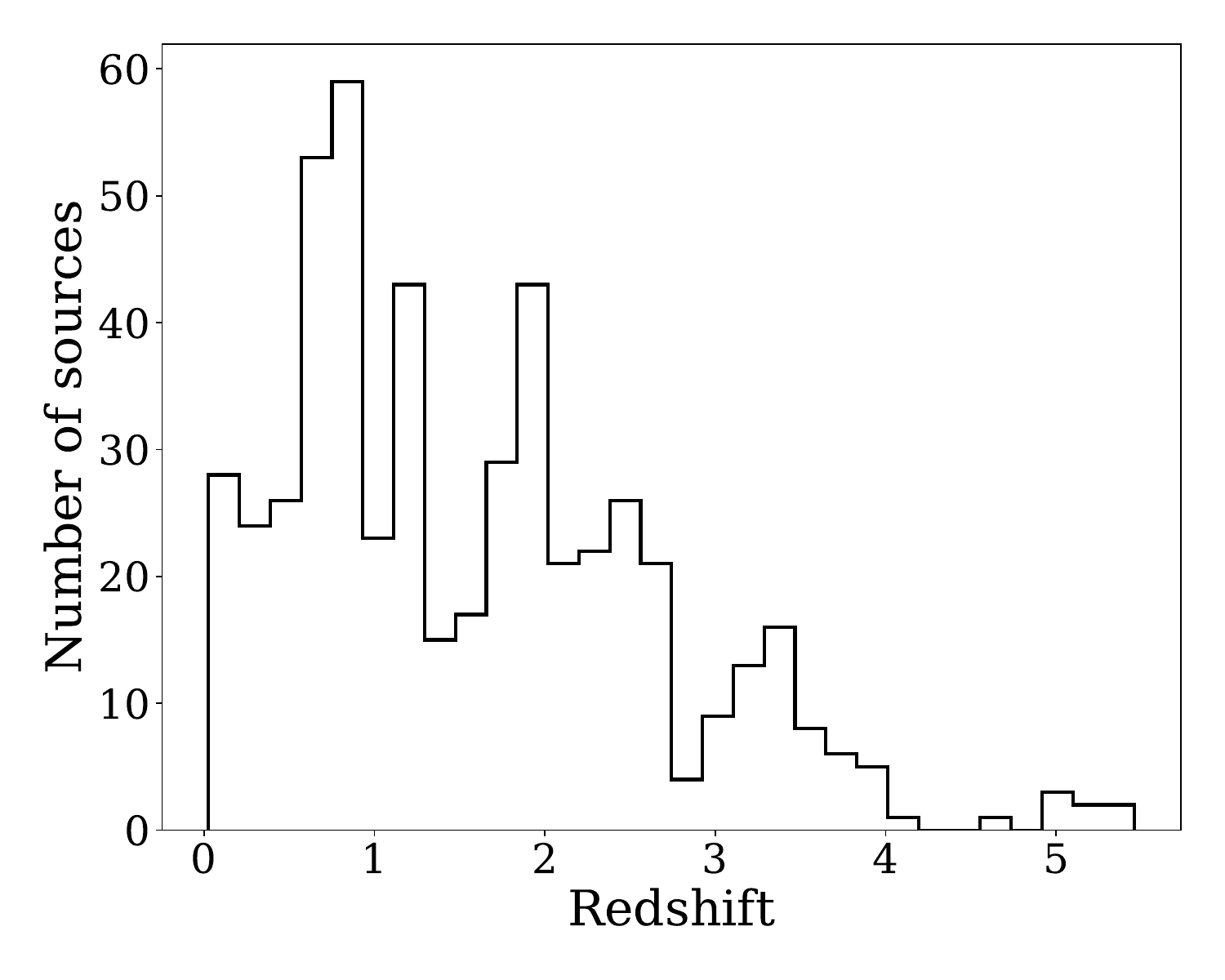}
    \caption{The redshift distribution of our COSMOS2020 matched MIRI \SI{5.6}{\micro \meter}m sources. These are LePhare redshifts from the classic COSMOS2020 catalog.}
    \label{fig:redshifts}
\end{figure}

\subsection{The MIRI \SI{5.6}{\micro \meter} sources in the COSMOS2020 catalog \label{subsec:properties}}

As already discussed in Section\,\ref{sec:verification}, we cross-matched the sources within our two COSMOS fields with the COSMOS2020 catalog \citep{Weaver2022,Weaver2022_catalog}. The overlap between our MIRI detections, the COSMOS2020 sources and their subset with IRAC channel 3 detections is illustrated in Figure\,\ref{fig:crosscheck} in the case of the COSMOS1 field. Note that with these new data more than half of the COSMOS2020 sources now have \SI{5.6}{\micro \meter} detections. The number of sources with MIRI \SI{5.6}{\micro \meter} detection in this field is more than 4$\times$ the sources with previous IRAC channel 3 (\SI{5.8}{\micro \meter}) detections. This is illustrative of the fact that until {\sl JWST}, it was difficult to link the optical and mid-IR source populations.  Within our two COSMOS MIRI fields, there are 690 MIRI sources, of which 582 (84\%) are matched to COSMOS2020 catalog sources, using a 1$\arcsec$ matching radius (see Section\,\ref{sec:verification}).

Figure\,\ref{fig:redshifts} shows the redshift distribution of the MIRI sources with counterparts in the COSMOS2020 catalog. We show the LePhare redshifts; although the distribution is very similar using the EAZY redshifts. In Figure\,\ref{fig:redshifts}, we applied a quality cut of $0.01 < |z_{u68}-z_{l68}|/(1+z_{\mathrm{med}}) < 0.5$, where $u68$, $l68$, and $\mathrm{med}$ denote the upper and lower 68\% confidence intervals and the median of the redshift probability density function. Of the 582 matched sources, 520 passed the quality cut and are presented the figure. This histogram shows peaks at $z\sim1$ and $z\sim2$, i.e. we find many cosmic noon galaxies.

Up to $z\sim2.5$, detection is aided by a negative $k$-correction as the observed \SI{5.6}{\micro \meter} probes up the SED towards the stellar \SI{1.6}{\micro \meter} bump, which explains the sharper drop in sources at higher redshift. However, we do see a tail out to $z\sim5$ which corresponds to early massive galaxies.  

\begin{figure}[h]
    \centering
    \includegraphics[width = 0.95\linewidth]{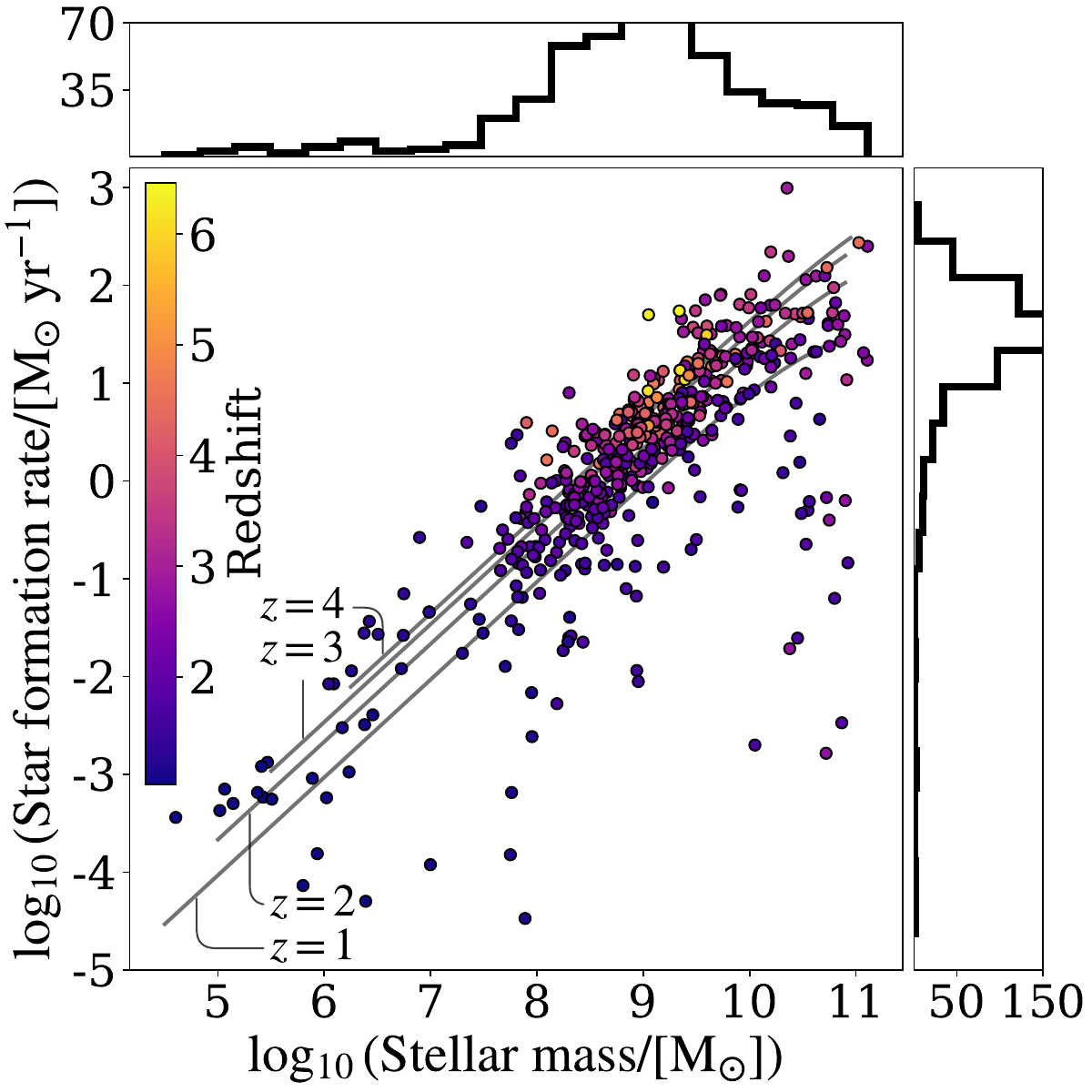}
    \caption{The stellar mass vs. star formation rate relation for the same sources presented in Figure\,\ref{fig:redshifts}. The overlaid lines are the main sequence relations at the indicated redshifts based on \citet{Koprowski2024}.}
    \label{fig:mass-sfr}
\end{figure}

Figure\,\ref{fig:mass-sfr} shows the stellar mass vs. star-formation rate for our COSMOS2020-matched MIRI sources. We find predominantly star-forming main sequence galaxies with stellar masses in the $10^8-10^{10}$\,M$_{\odot}$ range. We are probing the typical low-mass star-forming galaxy at cosmic noon, well below $M^*$ ($\approx10^{11}$M$_{\odot}$) at cosmic noon \citep{Adams2021}.

\subsection{The nature of MIRI sources not in COSMOS2020 \label{subsec:veryredmiri}}

As discussed in Section\,\ref{subsec:fakesources}, while the bulk of our sources (84\%) have counterparts in the COSMOS2020 catalog, 108 MIRI sources are un-matched. We examined these by eye to remove sources that have likely photometric issues such as being driven by a single hot pixel or being residuals of the stripes we saw prior to background subtraction (see Figure\,\ref{fig:bkg-sub}). After this visual inspection, we are left with 31 reliable MIRI detections not present in the COSMOS2020 catalog. This translates to a source density of $\sim$7 per square arcminute.

\newpage

\begin{figure}
    \centering
    \includegraphics[width = \linewidth]{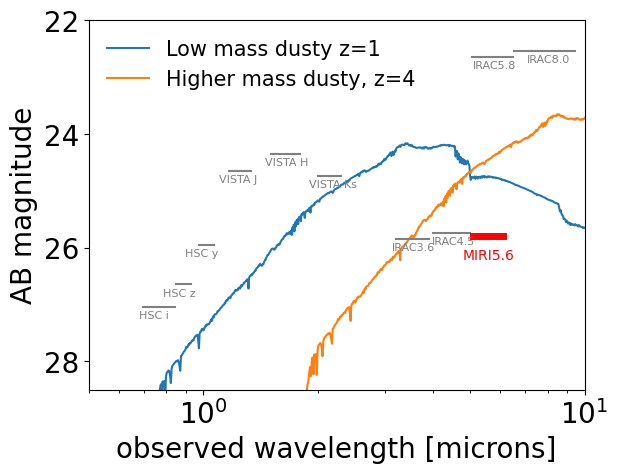}
    \caption{We compare the 5\,$\sigma$ magnitude limit of our images (in red) with the $izYJHK$+IRAC 5\,$\sigma$ magnitude limits for the COSMOS2020 catalog (in grey), based on \citet{Weaver2022}. We suggest two potential explanations for the MIRI detections that are not in the COSMOS catalog: either a low mass intermediate ($z\sim1$) redshift galaxy or a galaxy closer to $M_*$ at $z\sim3$ or beyond with high dust content. See Section 4.4 for more details on the models. }
    \label{fig:limiting_mags}
\end{figure}

To understand why strong MIRI detected sources may not be in the COSMOS2020 catalog consider that the COSMOS2020 source detection is based on a $\chi^2$-weighted $izYJHK$ image \citep{Weaver2022}. Figure\,\ref{fig:limiting_mags} shows the 5\,$\sigma$ limits of the COSMOS2020 catalog in these six bands, as well as the four IRAC channels. In all cases, we take the limits from the deep parts of COSMOS since these correspond to the locations of our two COSMOS fields. Note that our MIRI F560W depth is comparable to the IRAC channel 1 and 2 depths in this field, while the IRAC channel 3 image is approximately 3 magnitudes shallower (see figure\,\ref{fig:limiting_mags}). We overlay two potential models that could explain non-inclusion in the COSMOS2020 catalog, but detection in MIRI \SI{5.6}{\micro \meter}. One is a $z=1$ $M_{star}=10^9M\odot$ star-forming galaxy with $A_V=3$ (the ``Low-mass dusty'' model). The other is a $z=4$ $M_{star}=5\times10^{10}M\odot$ star-forming galaxy with $A_V=3$ (the ``High mass dusty'' model).  Here we use the Composite Stellar Population (CSP) models from \citet{Maraston2005}, assume an age of 300Myr for both galaxies and use the Calzetti law \citep{Calzetti1994} for dust attenuation. The first model is included because while, previously, low mass galaxies were typically not considered very dusty, {\sl JWST} has recently found evidence of a population of $z<2$ extremely dusty dwarfs (with masses $10^7-10^9M_{\odot}$ and up to $A_V\sim5$ , \citealt{Bisigello2023}; see also \citealt{Pope2023}).

\begin{figure*}[h!]
    \centering
    \includegraphics[width = 0.9\linewidth]{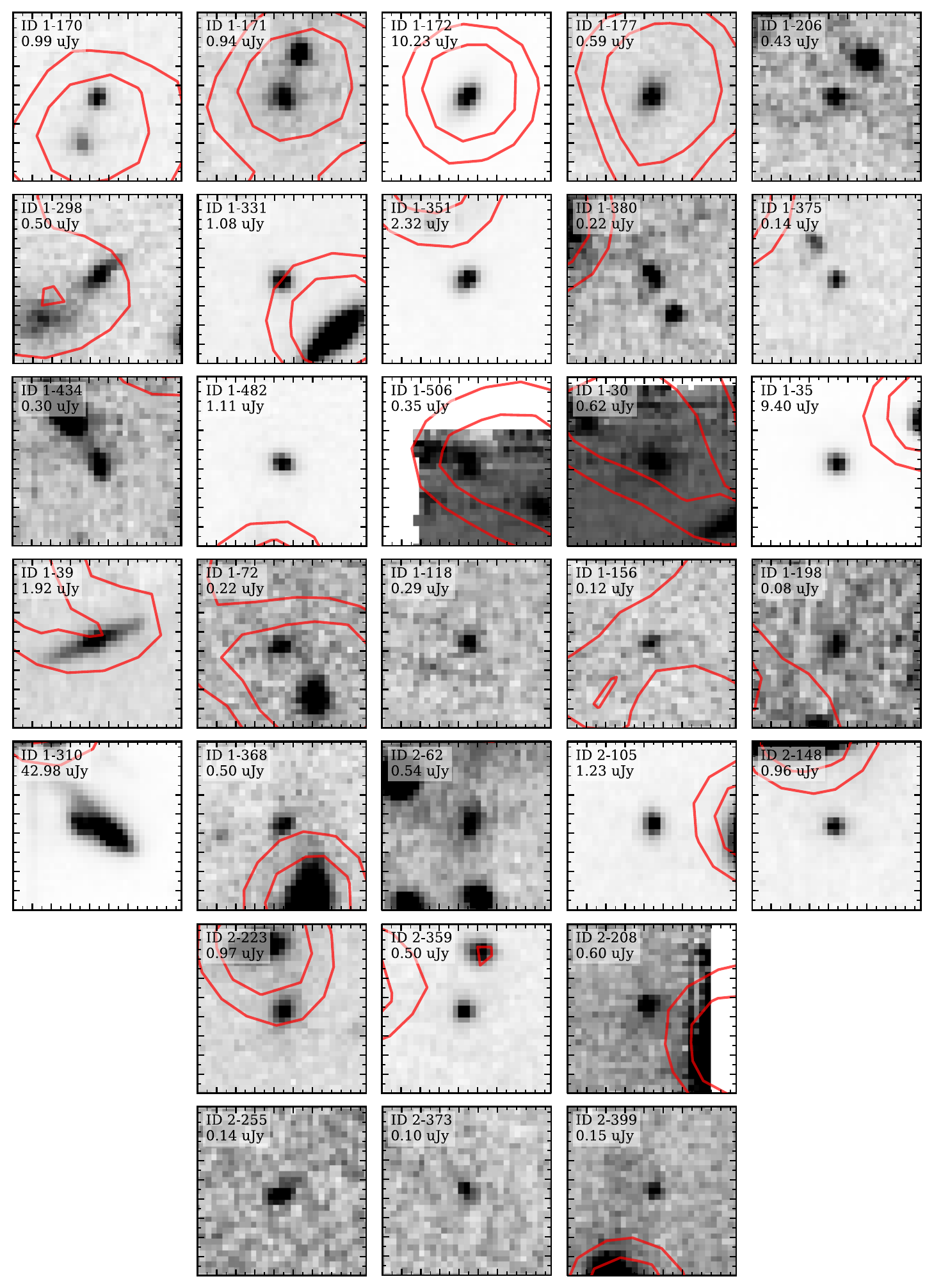}
    \caption{$3.3\times\ang{;;3.3}$ cutouts of the 31 sources detected in MIRI \SI{5.6}{\micro \meter} but are not present in the COSMOS 2020 Classic catalog. The red contours represent the 2 and 3\,$\sigma$ levels from the COSMOS IRAC channel 1 image from \citet{Annunziatella2023}, with an applied Gaussian blur ($\sigma = 0.8$). We label each source with our catalog ID and its MIRI \SI{5.6}{\micro \meter} flux.}
    \label{fig:miri_bright_cutouts}
\end{figure*}

In Figure\,\ref{fig:limiting_mags}, our model spectra are chosen to both lie below the limits for source detection in COSMOS2020 but also to have MIRI \SI{5.6}{\micro \meter} magnitudes on par with those of the sources shown in Figure\,\ref{fig:miri_bright_cutouts}. Both models satisfy these criteria, but they are clearly distinguishable by the IRAC channel 1 brightness. Therefore in Figure\,\ref{fig:miri_bright_cutouts} we overlay our MIRI images of these 31 sources with the 2 and 3\,$\sigma$ contours from the COSMOS IRAC channel 1 image. We used the IRAC channel 1 mosaic combining all available COSMOS data, as produced by \citet{Annunziatella2023}. We find that 3-7 of the sources have emission in IRAC channel 1. The range given is to distinguish clean detections from ones affected by blending in the IRAC image. The three clean IRAC detections in particular (ID 1-16, 1-107, 1-110) are all consistent with the low-mass dusty model. However, the bulk of our 31 sources are undetected in these IRAC images suggesting they are higher-mass and higher redshift dusty sources. Further study of these 31 sources, folding on the new COSMOS-Web data \citep{Casey2023}, as well as all other available data in COSMOS is reserved for a separate paper.

\section{Summary \& Conclusions}

In this paper, we present a study of the number counts and source properties of sources detected in eight MIRI F560W images with a combined area of 18.4\,sq.arcmin and a 5\,$\sigma$ depth of $\approx \SI{0.1}{\micro Jy}$. Two of our eight fields overlap with the COSMOS2020 catalog which also allows for a more detailed look at the source population properties. Below we summarize our key findings:
\begin{itemize}
\item The pipeline reduced MIRI F560W images required further background subtraction, given strong striping artifacts. We also note some residual structure seen in the noise, consistent with tree ring artifacts as noted in $JWST$ documentation. The photometric zero point needed to be adjusted since the header keyword in these early images was not using the true measured MIRI PSF.

\item  Our study includes 8 widely separated fields, allowing us to explore the effects of cosmic variance. We used this to construct field-to-field variation uncertainty on our combined MIRI \SI{5.6}{\micro \meter} number counts.

\item Our number counts have a more pronounced knee, at $\approx \SI{2}{\micro Jy}$ and are $\approx2\times$ higher than those computed from the CEERS images \citep{Yang2023}. This difference however is only at $\approx$1\,$\sigma$ level given our measured field-to-field variation. The observed counts are consistent with the \citet{cowley:2018} SAM predictions around the knee. These models however overpredict the observed counts below the knee of the counts. 

\item We find 84\% of the MIRI sources in the COSMOS fields have counterparts in the COSMOS2020 catalog. They are predominantly at cosmic noon, and often dusty star-forming galaxies with stellar masses well below $M^*$.

\item We find 31 very red sources that are 
reliably detected in MIRI, but are not in the COSMOS2020 catalog. This population has a source density of $\approx$7 per square arcminute. They are consistent with being either intermediate redshift ($z\sim1$) dusty lower mass galaxies (those with a strong IRAC channel 1 detection, 10 to 20\%) or high mass, high-$z$ ($z\gtrsim4$) dusty galaxies (those with weak or no IRAC channel 1 detections, 80 to 90\%). These MIRI \SI{5.6}{\micro \meter} only sources will be explored further in a follow-up paper.  

\end{itemize}

All the {\it JWST} data used in this paper can be found in MAST: \dataset[10.17909/jsqw-mq02]{http://dx.doi.org/10.17909/jsqw-mq02}.

\section*{Acknowledgments}

We are grateful to the anonymous referee for their careful reading of our manuscript and detailed feedback which greatly improved the clarity of this paper.  We are grateful to Meredith Stone for providing us with her MIRI 5.6\,$\mu$m counts ahead of publication. We are also grateful to Matthew Ashby for providing us with a copy of his old IRAC channel 3 counts. Based on observations with the NASA/ESA/CSA {\sl JWST} obtained at the Space Telescope Science Institute, which is operated by the Association of Universities for Research in Astronomy, Incorporated, under NASA contract NAS5-03127. Support for program number JWST-GO-01762 was provided through a grant from the STScI under grant number JWST-GO-01762.010-A.

\bibliography{counts_paper_refs}{}

\begin{thebibliography}{}
\expandafter\ifx\csname natexlab\endcsname\relax\def\natexlab#1{#1}\fi
\providecommand{\url}[1]{\href{#1}{#1}}
\providecommand{\dodoi}[1]{doi:~\href{http://doi.org/#1}{\nolinkurl{#1}}}
\providecommand{\doeprint}[1]{\href{http://ascl.net/#1}{\nolinkurl{http://ascl.net/#1}}}
\providecommand{\doarXiv}[1]{\href{https://arxiv.org/abs/#1}{\nolinkurl{https://arxiv.org/abs/#1}}}

\bibitem[{{Adams} {et~al.}(2021){Adams}, {Bowler}, {Jarvis}, {H{\"a}u{\ss}ler},
  \& {Lagos}}]{Adams2021}
{Adams}, N.~J., {Bowler}, R.~A.~A., {Jarvis}, M.~J., {H{\"a}u{\ss}ler}, B., \&
  {Lagos}, C.~D.~P. 2021, \mnras, 506, 4933, \dodoi{10.1093/mnras/stab1956}

\bibitem[{{Alberts} {et~al.}(2024){Alberts}, {Lyu}, {Shivaei}, {Rieke},
  {Perez-Gonzalez}, {Bonventura}, {Zhu}, {Helton}, {Ji}, {Morrison},
  {Robertson}, {Stone}, {Sun}, {Williams}, \& {Willmer}}]{Alberts2024}
{Alberts}, S., {Lyu}, J., {Shivaei}, I., {et~al.} 2024, arXiv e-prints,
  arXiv:2405.15972.
\newblock \doarXiv{2405.15972}

\bibitem[{{Annunziatella} {et~al.}(2023){Annunziatella}, {Sajina}, {Stefanon},
  {Marchesini}, {Lacy}, {Labb{\'e}}, {Houston}, {Bezanson}, {Egami}, {Fan},
  {Farrah}, {Greene}, {Goulding}, {Lin}, {Liu}, {Moutard}, {Ono}, {Ouchi},
  {Sawicki}, {Surace}, \& {Whitaker}}]{Annunziatella2023}
{Annunziatella}, M., {Sajina}, A., {Stefanon}, M., {et~al.} 2023, \aj, 166, 25,
  \dodoi{10.3847/1538-3881/acd773}

\bibitem[{{Ashby} {et~al.}(2009){Ashby}, {Stern}, {Brodwin}, {Griffith},
  {Eisenhardt}, {Koz{\l}owski}, {Kochanek}, {Bock}, {Borys}, {Brand}, {Brown},
  {Cool}, {Cooray}, {Croft}, {Dey}, {Eisenstein}, {Gonzalez}, {Gorjian},
  {Grogin}, {Ivison}, {Jacob}, {Jannuzi}, {Mainzer}, {Moustakas},
  {R{\"o}ttgering}, {Seymour}, {Smith}, {Stanford}, {Stauffer}, {Sullivan},
  {van Breugel}, {Willner}, \& {Wright}}]{Ashby2009}
{Ashby}, M.~L.~N., {Stern}, D., {Brodwin}, M., {et~al.} 2009, \apj, 701, 428,
  \dodoi{10.1088/0004-637X/701/1/428}

\bibitem[{{Barrufet} {et~al.}(2023){Barrufet}, {Oesch}, {Weibel}, {Brammer},
  {Bezanson}, {Bouwens}, {Fudamoto}, {Gonzalez}, {Gottumukkala}, {Illingworth},
  {Heintz}, {Holden}, {Labbe}, {Magee}, {Naidu}, {Nelson}, {Stefanon}, {Smit},
  {van Dokkum}, {Weaver}, \& {Williams}}]{Barrufet2023}
{Barrufet}, L., {Oesch}, P.~A., {Weibel}, A., {et~al.} 2023, \mnras, 522, 449,
  \dodoi{10.1093/mnras/stad947}

\bibitem[{{Bertin, E.} \& {Arnouts, S.}(1996)}]{Bertin:1996}
{Bertin, E.}, \& {Arnouts, S.} 1996, Astron. Astrophys. Suppl. Ser., 117, 393,
  \dodoi{10.1051/aas:1996164}

\bibitem[{{Bisigello} {et~al.}(2023){Bisigello}, {Gandolfi}, {Grazian},
  {Rodighiero}, {Costantin}, {Cooray}, {Feltre}, {Gruppioni}, {Hathi},
  {Holwerda}, {Koekemoer}, {Lucas}, {Newman}, {P{\'e}rez-Gonz{\'a}lez}, {Yung},
  {de la Vega}, {Arrabal Haro}, {Bagley}, {Dickinson}, {Finkelstein},
  {Kartaltepe}, {Papovich}, {Pirzkal}, \& {Wilkins}}]{Bisigello2023}
{Bisigello}, L., {Gandolfi}, G., {Grazian}, A., {et~al.} 2023, \aap, 676, A76,
  \dodoi{10.1051/0004-6361/202346219}

\bibitem[{{Brammer} {et~al.}(2008){Brammer}, {van Dokkum}, \&
  {Coppi}}]{Brammer2008}
{Brammer}, G.~B., {van Dokkum}, P.~G., \& {Coppi}, P. 2008, \apj, 686, 1503,
  \dodoi{10.1086/591786}

\bibitem[{{Calzetti} {et~al.}(1994){Calzetti}, {Kinney}, \&
  {Storchi-Bergmann}}]{Calzetti1994}
{Calzetti}, D., {Kinney}, A.~L., \& {Storchi-Bergmann}, T. 1994, \apj, 429,
  582, \dodoi{10.1086/174346}

\bibitem[{{Casey} {et~al.}(2014){Casey}, {Narayanan}, \& {Cooray}}]{Casey2014}
{Casey}, C.~M., {Narayanan}, D., \& {Cooray}, A. 2014, \physrep, 541, 45,
  \dodoi{10.1016/j.physrep.2014.02.009}

\bibitem[{{Casey} {et~al.}(2023){Casey}, {Kartaltepe}, {Drakos}, {Franco},
  {Harish}, {Paquereau}, {Ilbert}, {Rose}, {Cox}, {Nightingale}, {Robertson},
  {Silverman}, {Koekemoer}, {Massey}, {McCracken}, {Rhodes}, {Akins}, {Allen},
  {Amvrosiadis}, {Arango-Toro}, {Bagley}, {Bongiorno}, {Capak}, {Champagne},
  {Chartab}, {Ch{\'a}vez Ortiz}, {Chworowsky}, {Cooke}, {Cooper}, {Darvish},
  {Ding}, {Faisst}, {Finkelstein}, {Fujimoto}, {Gentile}, {Gillman}, {Gould},
  {Gozaliasl}, {Hayward}, {He}, {Hemmati}, {Hirschmann}, {Jahnke}, {Jin},
  {Khostovan}, {Kokorev}, {Lambrides}, {Laigle}, {Larson}, {Leung}, {Liu},
  {Liaudat}, {Long}, {Magdis}, {Mahler}, {Mainieri}, {Manning}, {Maraston},
  {Martin}, {McCleary}, {McKinney}, {McPartland}, {Mobasher}, {Pattnaik},
  {Renzini}, {Rich}, {Sanders}, {Sattari}, {Scognamiglio}, {Scoville}, {Sheth},
  {Shuntov}, {Sparre}, {Suzuki}, {Talia}, {Toft}, {Trakhtenbrot}, {Urry},
  {Valentino}, {Vanderhoof}, {Vardoulaki}, {Weaver}, {Whitaker}, {Wilkins},
  {Yang}, \& {Zavala}}]{Casey2023}
{Casey}, C.~M., {Kartaltepe}, J.~S., {Drakos}, N.~E., {et~al.} 2023, \apj, 954,
  31, \dodoi{10.3847/1538-4357/acc2bc}

\bibitem[{{Cowley} {et~al.}(2018){Cowley}, {Baugh}, {Cole}, {Frenk}, \&
  {Lacey}}]{cowley:2018}
{Cowley}, W.~I., {Baugh}, C.~M., {Cole}, S., {Frenk}, C.~S., \& {Lacey}, C.~G.
  2018, \mnras, 474, 2352, \dodoi{10.1093/mnras/stx2897}

\bibitem[{{Elsner} {et~al.}(2008){Elsner}, {Feulner}, \& {Hopp}}]{Elsner2008}
{Elsner}, F., {Feulner}, G., \& {Hopp}, U. 2008, \aap, 477, 503,
  \dodoi{10.1051/0004-6361:20078343}

\bibitem[{Finkelstein {et~al.}(2022)Finkelstein, Bagley, Haro, Dickinson,
  Ferguson, Kartaltepe, Papovich, Burgarella, Kocevski, Iyer,
  {et~al.}}]{finkelstein2022long}
Finkelstein, S.~L., Bagley, M.~B., Haro, P.~A., {et~al.} 2022, The
  Astrophysical journal letters, 940, L55

\bibitem[{{Gardner} {et~al.}(2006){Gardner}, {Mather}, {Clampin}, {Doyon},
  {Greenhouse}, {Hammel}, {Hutchings}, {Jakobsen}, {Lilly}, {Long}, {Lunine},
  {McCaughrean}, {Mountain}, {Nella}, {Rieke}, {Rieke}, {Rix}, {Smith},
  {Sonneborn}, {Stiavelli}, {Stockman}, {Windhorst}, \& {Wright}}]{Gardner2006}
{Gardner}, J.~P., {Mather}, J.~C., {Clampin}, M., {et~al.} 2006, \ssr, 123,
  485, \dodoi{10.1007/s11214-006-8315-7}

\bibitem[{{G{\'a}sp{\'a}r} {et~al.}(2021){G{\'a}sp{\'a}r}, {Rieke}, {Guillard},
  {Dicken}, {Gastaud}, {Alberts}, {Morrison}, {Ressler}, {Argyriou}, \&
  {Glasse}}]{Gaspar2021}
{G{\'a}sp{\'a}r}, A., {Rieke}, G.~H., {Guillard}, P., {et~al.} 2021, \pasp,
  133, 014504, \dodoi{10.1088/1538-3873/abcd04}

\bibitem[{{Hughes} {et~al.}(1998){Hughes}, {Serjeant}, {Dunlop},
  {Rowan-Robinson}, {Blain}, {Mann}, {Ivison}, {Peacock}, {Efstathiou}, {Gear},
  {Oliver}, {Lawrence}, {Longair}, {Goldschmidt}, \& {Jenness}}]{Hughes1998}
{Hughes}, D.~H., {Serjeant}, S., {Dunlop}, J., {et~al.} 1998, \nat, 394, 241,
  \dodoi{10.1038/28328}

\bibitem[{{Kim} {et~al.}(2024){Kim}, {Goto}, {Ling}, {Wu}, {Hashimoto},
  {Kilerci}, {Ho}, {Uno}, {Wang}, \& {Lin}}]{Kim2024}
{Kim}, S.~J., {Goto}, T., {Ling}, C.-T., {et~al.} 2024, \mnras, 527, 5525,
  \dodoi{10.1093/mnras/stad3499}

\bibitem[{{Kirkpatrick} {et~al.}(2023){Kirkpatrick}, {Yang}, {Le Bail},
  {Troiani}, {Bell}, {Cleri}, {Elbaz}, {Finkelstein}, {Hathi}, {Hirschmann},
  {Holwerda}, {Kocevski}, {Lucas}, {McKinney}, {Papovich}, {Perez-Gonzalez},
  {de la Vega}, {Bagley}, {Daddi}, {Dickinson}, {Ferguson}, {Fontana},
  {Grazian}, {Grogin}, {Arrabal Haro}, {Kartaltepe}, {Kewley}, {Koekemoer},
  {Lotz}, {Pentericci}, {Pirzkal}, {Ravindranath}, {Somerville}, {Trump},
  {Wilkins}, \& {Yung}}]{Kirkpatrick2023}
{Kirkpatrick}, A., {Yang}, G., {Le Bail}, A., {et~al.} 2023, arXiv e-prints,
  arXiv:2308.09750, \dodoi{10.48550/arXiv.2308.09750}

\bibitem[{{Koprowski} {et~al.}(2024){Koprowski}, {Wijesekera}, {Dunlop},
  {McLeod}, {Micha{\l}owski}, {Lisiecki}, \& {McLure}}]{Koprowski2024}
{Koprowski}, M.~P., {Wijesekera}, J.~V., {Dunlop}, J.~S., {et~al.} 2024, arXiv
  e-prints, arXiv:2403.06575, \dodoi{10.48550/arXiv.2403.06575}

\bibitem[{{La Torre} {et~al.}(2024){La Torre}, {Sajina}, {Goulding},
  {Marchesini}, {Bezanson}, {Pearl}, \& {Sodr{\'e}}}]{LaTorre2024}
{La Torre}, V., {Sajina}, A., {Goulding}, A.~D., {et~al.} 2024, \aj, 167, 261,
  \dodoi{10.3847/1538-3881/ad3821}

\bibitem[{{Labbe} {et~al.}(2023){Labbe}, {Greene}, {Bezanson}, {Fujimoto},
  {Furtak}, {Goulding}, {Matthee}, {Naidu}, {Oesch}, {Atek}, {Brammer},
  {Chemerynska}, {Coe}, {Cutler}, {Dayal}, {Feldmann}, {Franx}, {Glazebrook},
  {Leja}, {Marchesini}, {Maseda}, {Nanayakkara}, {Nelson}, {Pan}, {Papovich},
  {Price}, {Suess}, {Wang}, {Whitaker}, {Williams}, \& {Zitrin}}]{Labbe2023}
{Labbe}, I., {Greene}, J.~E., {Bezanson}, R., {et~al.} 2023, arXiv e-prints,
  arXiv:2306.07320, \dodoi{10.48550/arXiv.2306.07320}

\bibitem[{{Lacey} {et~al.}(2016){Lacey}, {Baugh}, {Frenk}, {Benson}, {Bower},
  {Cole}, {Gonzalez-Perez}, {Helly}, {Lagos}, \& {Mitchell}}]{Lacey2016}
{Lacey}, C.~G., {Baugh}, C.~M., {Frenk}, C.~S., {et~al.} 2016, \mnras, 462,
  3854, \dodoi{10.1093/mnras/stw1888}

\bibitem[{{Lacy} {et~al.}(2021){Lacy}, {Surace}, {Farrah}, {Nyland}, {Afonso},
  {Brandt}, {Clements}, {Lagos}, {Maraston}, {Pforr}, {Sajina}, {Sako},
  {Vaccari}, {Wilson}, {Ballantyne}, {Barkhouse}, {Brunner}, {Cane}, {Clarke},
  {Cooper}, {Cooray}, {Covone}, {D'Andrea}, {Evrard}, {Ferguson}, {Frieman},
  {Gonzalez-Perez}, {Gupta}, {Hatziminaoglou}, {Huang}, {Jagannathan},
  {Jarvis}, {Jones}, {Kimball}, {Lidman}, {Lubin}, {Marchetti}, {Martini},
  {McMahon}, {Mei}, {Messias}, {Murphy}, {Newman}, {Nichol}, {Norris},
  {Oliver}, {Perez-Fournon}, {Peters}, {Pierre}, {Polisensky}, {Richards},
  {Ridgway}, {R{\"o}ttgering}, {Seymour}, {Shirley}, {Somerville}, {Strauss},
  {Suntzeff}, {Thorman}, {van Kampen}, {Verma}, {Wechsler}, \&
  {Wood-Vasey}}]{Lacy2021}
{Lacy}, M., {Surace}, J.~A., {Farrah}, D., {et~al.} 2021, \mnras, 501, 892,
  \dodoi{10.1093/mnras/staa3714}

\bibitem[{{Ling} {et~al.}(2022){Ling}, {Kim}, {Wu}, {Goto}, {Kilerci},
  {Hashimoto}, {Lin}, {Wang}, {Ho}, \& {Hsiao}}]{Ling2022}
{Ling}, C.-T., {Kim}, S.~J., {Wu}, C. K.~W., {et~al.} 2022, \mnras, 517, 853,
  \dodoi{10.1093/mnras/stac2716}

\bibitem[{{Lyu} {et~al.}(2024){Lyu}, {Alberts}, {Rieke}, {Shivaei},
  {P{\'e}rez-Gonz{\'a}lez}, {Sun}, {Hainline}, {Baum}, {Bonaventura}, {Bunker},
  {Egami}, {Eisenstein}, {Florian}, {Ji}, {Johnson}, {Morrison}, {Rieke},
  {Robertson}, {Rujopakarn}, {Tacchella}, {Scholtz}, \& {Willmer}}]{Lyu2024}
{Lyu}, J., {Alberts}, S., {Rieke}, G.~H., {et~al.} 2024, \apj, 966, 229,
  \dodoi{10.3847/1538-4357/ad3643}

\bibitem[{{Madau} \& {Dickinson}(2014)}]{MadauDickinson2014}
{Madau}, P., \& {Dickinson}, M. 2014, \araa, 52, 415,
  \dodoi{10.1146/annurev-astro-081811-125615}

\bibitem[{{Maraston}(2005)}]{Maraston2005}
{Maraston}, C. 2005, \mnras, 362, 799, \dodoi{10.1111/j.1365-2966.2005.09270.x}

\bibitem[{{Martis} {et~al.}(2023){Martis}, {Marchesini}, {Muzzin}, {Willott},
  \& {Sawicki}}]{martis2023}
{Martis}, N.~S., {Marchesini}, D.~M., {Muzzin}, A., {Willott}, C.~J., \&
  {Sawicki}, M. 2023, \mnras, 518, 4961, \dodoi{10.1093/mnras/stac3455}

\bibitem[{{Muzzin} {et~al.}(2009){Muzzin}, {Marchesini}, {van Dokkum},
  {Labb{\'e}}, {Kriek}, \& {Franx}}]{Muzzin2009}
{Muzzin}, A., {Marchesini}, D., {van Dokkum}, P.~G., {et~al.} 2009, \apj, 701,
  1839, \dodoi{10.1088/0004-637X/701/2/1839}

\bibitem[{{Oke} \& {Gunn}(1974)}]{OkeGunn1974}
{Oke}, J.~B., \& {Gunn}, J.~E. 1974, \apjl, 189, L5, \dodoi{10.1086/181450}

\bibitem[{{Papovich} {et~al.}(2023){Papovich}, {Cole}, {Yang}, {Finkelstein},
  {Barro}, {Buat}, {Burgarella}, {P{\'e}rez-Gonz{\'a}lez}, {Santini},
  {Seill{\'e}}, {Shen}, {Arrabal Haro}, {Bagley}, {Bell}, {Bisigello},
  {Calabr{\`o}}, {Casey}, {Castellano}, {Chworowsky}, {Cleri}, {Costantin},
  {Cooper}, {Dickinson}, {Ferguson}, {Fontana}, {Giavalisco}, {Grazian},
  {Grogin}, {Hathi}, {Holwerda}, {Hutchison}, {Kartaltepe}, {Kewley},
  {Kirkpatrick}, {Kocevski}, {Koekemoer}, {Larson}, {Long}, {Lucas},
  {Pentericci}, {Pirzkal}, {Ravindranath}, {Somerville}, {Trump}, {Urbano
  Stawinski}, {Weiner}, {Wilkins}, {Yung}, \& {Zavala}}]{papovich:2023}
{Papovich}, C., {Cole}, J.~W., {Yang}, G., {et~al.} 2023, \apjl, 949, L18,
  \dodoi{10.3847/2041-8213/acc948}

\bibitem[{{P{\'e}rez-Gonz{\'a}lez} {et~al.}(2024){P{\'e}rez-Gonz{\'a}lez},
  {Barro}, {Rieke}, {Lyu}, {Rieke}, {Alberts}, {Williams}, {Hainline}, {Sun},
  {Puskas}, {Annunziatella}, {Baker}, {Bunker}, {Egami}, {Ji}, {Johnson},
  {Robertson}, {Rodriguez Del Pino}, {Rujopakarn}, {Shivaei}, {Tacchella},
  {Willmer}, \& {Willott}}]{Perez-Gonzalez2024}
{P{\'e}rez-Gonz{\'a}lez}, P.~G., {Barro}, G., {Rieke}, G.~H., {et~al.} 2024,
  arXiv e-prints, arXiv:2401.08782, \dodoi{10.48550/arXiv.2401.08782}

\bibitem[{{Pope} {et~al.}(2008){Pope}, {Bussmann}, {Dey}, {Meger}, {Alexander},
  {Brodwin}, {Chary}, {Dickinson}, {Frayer}, {Greve}, {Huynh}, {Lin},
  {Morrison}, {Scott}, \& {Yan}}]{Pope2008}
{Pope}, A., {Bussmann}, R.~S., {Dey}, A., {et~al.} 2008, \apj, 689, 127,
  \dodoi{10.1086/592739}

\bibitem[{{Pope} {et~al.}(2021){Pope}, {Sajina}, {Yan}, {Alberts}, {Armus},
  {Bonato}, {Chary}, {Coppin}, {Dale}, {Farrah}, {Ferkinhoff}, {Groves},
  {Hayward}, {Kirkpatrick}, {Lagache}, {McKinney}, {Murphy}, {Nesvadba},
  {Ogle}, \& {Veilleux}}]{halfway:2021}
{Pope}, A., {Sajina}, A., {Yan}, L., {et~al.} 2021, {Halfway to the peak: A
  bridge program to map coeval star formation and supermassive black hole
  growth}, JWST Proposal. Cycle 1, ID. \#1762

\bibitem[{{Pope} {et~al.}(2023){Pope}, {McKinney}, {Kamieneski}, {Battisti},
  {Aretxaga}, {Brammer}, {Diego}, {Hughes}, {Keller}, {Marchesini}, {Mizener},
  {Monta{\~n}a}, {Murphy}, {Whitaker}, {Wilson}, \& {Yun}}]{Pope2023}
{Pope}, A., {McKinney}, J., {Kamieneski}, P., {et~al.} 2023, \apjl, 951, L46,
  \dodoi{10.3847/2041-8213/acdf5a}

\bibitem[{{Puget} {et~al.}(1996){Puget}, {Abergel}, {Bernard}, {Boulanger},
  {Burton}, {Desert}, \& {Hartmann}}]{Puget1996}
{Puget}, J.~L., {Abergel}, A., {Bernard}, J.~P., {et~al.} 1996, \aap, 308, L5

\bibitem[{{Rieke} {et~al.}(2015){Rieke}, {Wright}, {B{\"o}ker}, {Bouwman},
  {Colina}, {Glasse}, {Gordon}, {Greene}, {G{\"u}del}, {Henning}, {Justtanont},
  {Lagage}, {Meixner}, {N{\o}rgaard-Nielsen}, {Ray}, {Ressler}, {van Dishoeck},
  \& {Waelkens}}]{Rieke2015}
{Rieke}, G.~H., {Wright}, G.~S., {B{\"o}ker}, T., {et~al.} 2015, \pasp, 127,
  584, \dodoi{10.1086/682252}

\bibitem[{{Rigby} {et~al.}(2023){Rigby}, {Perrin}, {McElwain}, {Kimble},
  {Friedman}, {Lallo}, {Doyon}, {Feinberg}, {Ferruit}, {Glasse}, {Rieke},
  {Rieke}, {Wright}, {Willott}, {Colon}, {Milam}, {Neff}, {Stark}, {Valenti},
  {Abell}, {Abney}, {Abul-Huda}, {Acton}, {Adams}, {Adler}, {Aguilar}, {Ahmed},
  {Albert}, {Alberts}, {Aldridge}, {Allen}, {Altenburg},
  {{\'A}lvarez-M{\'a}rquez}, {Alves de Oliveira}, {Andersen}, {Anderson},
  {Anderson}, {Argyriou}, {Armstrong}, {Arribas}, {Artigau}, {Arvai},
  {Atkinson}, {Bacon}, {Bair}, {Banks}, {Barrientes}, {Barringer}, {Bartosik},
  {Bast}, {Baudoz}, {Beatty}, {Bechtold}, {Beck}, {Bergeron}, {Bergkoetter},
  {Bhatawdekar}, {Birkmann}, {Blazek}, {Blome}, {Boccaletti}, {B{\"o}ker},
  {Boia}, {Bonaventura}, {Bond}, {Bosley}, {Boucarut}, {Bourque}, {Bouwman},
  {Bower}, {Bowers}, {Boyer}, {Bradley}, {Brady}, {Braun}, {Breda},
  {Bresnahan}, {Bright}, {Britt}, {Bromenschenkel}, {Brooks}, {Brooks},
  {Brown}, {Brown}, {Brown}, {Bunker}, {Burger}, {Bushouse}, {Cale}, {Cameron},
  {Cameron}, {Canipe}, {Caplinger}, {Caputo}, {Cara}, {Carey}, {Carniani},
  {Carrasquilla}, {Carruthers}, {Case}, {Catherine}, {Chance}, {Chapman},
  {Charlot}, {Charlow}, {Chayer}, {Chen}, {Cherinka}, {Chichester}, {Chilton},
  {Chonis}, {Clampin}, {Clark}, {Clark}, {Coe}, {Coleman}, {Comber}, {Comeau},
  {Connolly}, {Cooper}, {Cooper}, {Coppock}, {Correnti}, {Cossou}, {Coulais},
  {Coyle}, {Cracraft}, {Curti}, {Cuturic}, {Davis}, {Davis}, {Dean}, {DeLisa},
  {deMeester}, {Dencheva}, {Dencheva}, {DePasquale}, {Deschenes}, {Hunor
  Detre}, {Diaz}, {Dicken}, {DiFelice}, {Dillman}, {Dixon}, {Doggett},
  {Donaldson}, {Douglas}, {DuPrie}, {Dupuis}, {Durning}, {Easmin}, {Eck},
  {Edeani}, {Egami}, {Ehrenwinkler}, {Eisenhamer}, {Eisenhower}, {Elie},
  {Elliott}, {Elliott}, {Ellis}, {Engesser}, {Espinoza}, {Etienne}, {Etxaluze},
  {Falini}, {Feeney}, {Ferry}, {Filippazzo}, {Fincham}, {Fix}, {Flagey},
  {Florian}, {Flynn}, {Fontanella}, {Ford}, {Forshay}, {Fox}, {Franz}, {Fu},
  {Fullerton}, {Galkin}, {Galyer}, {Garc{\'\i}a Mar{\'\i}n}, {Gardner},
  {Gardner}, {Garland}, {Garrett}, {Gasman}, {Gaspar}, {Gaudreau}, {Gauthier},
  {Geers}, {Geithner}, {Gennaro}, {Giardino}, {Girard}, {Giuliano},
  {Glassmire}, {Glauser}, {Glazer}, {Godfrey}, {Golimowski}, {Gollnitz},
  {Gong}, {Gonzaga}, {Gordon}, {Gordon}, {Goudfrooij}, {Greene}, {Greenhouse},
  {Grimaldi}, {Groebner}, {Grundy}, {Guillard}, {Gutman}, {Ha}, {Haderlein},
  {Hagedorn}, {Hainline}, {Haley}, {Hami}, {Hamilton}, {Hammel}, {Hansen},
  {Harkins}, {Harr}, {Hart}, {Hart}, {Hartig}, {Hashimoto}, {Haskins},
  {Hathaway}, {Havey}, {Hayden}, {Hecht}, {Heller-Boyer}, {Henriques}, {Henry},
  {Hermann}, {Hernandez}, {Hesman}, {Hicks}, {Hilbert}, {Hines}, {Hoffman},
  {Holfeltz}, {Holler}, {Hoppa}, {Hott}, {Howard}, {Howard}, {Hunter},
  {Hunter}, {Hurst}, {Husemann}, {Hustak}, {Ilinca Ignat}, {Illingworth},
  {Irish}, {Jackson}, {Jahromi}, {Jakobsen}, {James}, {James}, {Januszewski},
  {Jenkins}, {Jirdeh}, {Johnson}, {Johnson}, {Jones}, {Jones}, {Jones},
  {Jones}, {Jordan}, {Jordan}, {Jurczyk}, {Jurling}, {Kaleida}, {Kalmanson},
  {Kammerer}, {Kang}, {Kao}, {Karakla}, {Kavanagh}, {Kelly}, {Kendrew},
  {Kennedy}, {Kenny}, {Keski-kuha}, {Keyes}, {Kidwell}, {Kinzel}, {Kirk},
  {Kirkpatrick}, {Kirshenblat}, {Klaassen}, {Knapp}, {Knight}, {Knollenberg},
  {Koehler}, {Koekemoer}, {Kovacs}, {Kulp}, {Kumari}, {Kyprianou}, {La Massa},
  {Labador}, {Labiano}, {Lagage}, {Lajoie}, {Lallo}, {Lam}, {Lamb}, {Lambros},
  {Lampenfield}, {Langston}, {Larson}, {Law}, {Lawrence}, {Lee}, {Leisenring},
  {Lepo}, {Leveille}, {Levenson}, {Levine}, {Levy}, {Lewis}, {Lewis},
  {Libralato}, {Lightsey}, {Link}, {Liu}, {Lo}, {Lockwood}, {Logue}, {Long},
  {Long}, {Loomis}, {Lopez-Caniego}, {Lorenzo Alvarez}, {Love-Pruitt}, {Lucy},
  {Luetzgendorf}, {Maghami}, {Maiolino}, {Major}, {Malla}, {Malumuth},
  {Manjavacas}, {Mannfolk}, {Marrione}, {Marston}, {Martel}, {Maschmann},
  {Masci}, {Masciarelli}, {Maszkiewicz}, {Mather}, {McKenzie}, {McLean},
  {McMaster}, {Melbourne}, {Mel{\'e}ndez}, {Menzel}, {Merz}, {Meyett}, {Meza},
  {Miskey}, {Misselt}, {Moller}, {Morrison}, {Morse}, {Moseley}, {Mosier},
  {Mountain}, {Mueckay}, {Mueller}, {Mullally}, {Murphy}, {Murray}, {Murray},
  {Mustelier}, {Muzerolle}, {Mycroft}, {Myers}, {Myrick}, {Nanavati}, {Nance},
  {Nayak}, {Naylor}, {Nelan}, {Nickson}, {Nielson}, {Nieto-Santisteban},
  {Nikolov}, {Noriega-Crespo}, {O'Shaughnessy}, {O'Sullivan}, {Ochs}, {Ogle},
  {Oleszczuk}, {Olmsted}, {Osborne}, {Ottens}, {Owens}, {Pacifici}, {Pagan},
  {Page}, {Park}, {Parrish}, {Patapis}, {Paul}, {Pauly}, {Pavlovsky}, {Pedder},
  {Peek}, {Pena-Guerrero}, {Penanen}, {Perez}, {Perna}, {Perriello},
  {Phillips}, {Pietraszkiewicz}, {Pinaud}, {Pirzkal}, {Pitman}, {Piwowar},
  {Platais}, {Player}, {Plesha}, {Pollizi}, {Polster}, {Pontoppidan},
  {Porterfield}, {Proffitt}, {Pueyo}, {Pulliam}, {Quirt}, {Quispe Neira},
  {Ramos Alarcon}, {Ramsay}, {Rapp}, {Rapp}, {Rauscher}, {Ravindranath},
  {Rawle}, {Regan}, {Reichard}, {Reis}, {Ressler}, {Rest}, {Reynolds}, {Rhue},
  {Richon}, {Rickman}, {Ridgaway}, {Ritchie}, {Rix}, {Robberto}, {Robinson},
  {Robinson}, {Robinson}, {Rock}, {Rodriguez}, {Rodriguez Del Pino}, {Roellig},
  {Rohrbach}, {Roman}, {Romelfanger}, {Rose}, {Roteliuk}, {Roth}, {Rothwell},
  {Rowlands}, {Roy}, {Royer}, {Royle}, {Rui}, {Rumler}, {Runnels}, {Russ},
  {Rustamkulov}, {Ryden}, {Ryer}, {Sabata}, {Sabatke}, {Sabbi}, {Samuelson},
  {Sapp}, {Sappington}, {Sargent}, {Sauer}, {Scheithauer}, {Schlawin},
  {Schlitz}, {Schmitz}, {Schneider}, {Schreiber}, {Schulze}, {Schwab}, {Scott},
  {Sembach}, {Shanahan}, {Shaughnessy}, {Shaw}, {Shawger}, {Shay}, {Sheehan},
  {Shen}, {Sherman}, {Shiao}, {Shih}, {Shivaei}, {Sienkiewicz}, {Sing},
  {Sirianni}, {Sivaramakrishnan}, {Skipper}, {Sloan}, {Slocum}, {Slowinski},
  {Smith}, {Smith}, {Smith}, {Smith}, {Snyder}, {Soh}, {Sohn}, {Soto},
  {Spencer}, {Stallcup}, {Stansberry}, {Starr}, {Starr}, {Stewart},
  {Stiavelli}, {Straughn}, {Strickland}, {Stys}, {Summers}, {Sun}, {Sunnquist},
  {Swade}, {Swam}, {Swaters}, {Swoish}, {Taylor}, {Taylor}, {Te Plate}, {Tea},
  {Teague}, {Telfer}, {Temim}, {Thatte}, {Thompson}, {Thompson}, {Thomson},
  {Tikkanen}, {Tippet}, {Todd}, {Toolan}, {Tran}, {Trejo}, {Truong},
  {Tsukamoto}, {Tustain}, {Tyra}, {Ubeda}, {Underwood}, {Uzzo}, {Van Campen},
  {Vandal}, {Vandenbussche}, {Vila}, {Volk}, {Wahlgren}, {Waldman}, {Walker},
  {Wander}, {Warfield}, {Warner}, {Wasiak}, {Watkins}, {Weaver}, {Weilert},
  {Weiser}, {Weiss}, {Weissman}, {Welty}, {West}, {Wheate}, {Wheatley},
  {Wheeler}, {White}, {Whiteaker}, {Whitehouse}, {Whiteleather}, {Whitman},
  {Williams}, {Willmer}, {Willoughby}, {Wilson}, {Wirth}, {Wislowski}, {Wolf},
  {Wolfe}, {Wolff}, {Workman}, {Wright}, {Wu}, {Wu}, {Wymer}, {Yates},
  {Yeager}, {Yeates}, {Yerger}, {Yoon}, {Young}, {Yu}, {Zak}, {Zeidler},
  {Zhou}, {Zielinski}, {Zincke}, \& {Zonak}}]{rigby:2023}
{Rigby}, J., {Perrin}, M., {McElwain}, M., {et~al.} 2023, \pasp, 135, 048001,
  \dodoi{10.1088/1538-3873/acb293}

\bibitem[{{Sajina} {et~al.}(2006){Sajina}, {Scott}, {Dennefeld}, {Dole},
  {Lacy}, \& {Lagache}}]{Sajina2006}
{Sajina}, A., {Scott}, D., {Dennefeld}, M., {et~al.} 2006, \mnras, 369, 939,
  \dodoi{10.1111/j.1365-2966.2006.10361.x}

\bibitem[{{Sanders} {et~al.}(2007){Sanders}, {Salvato}, {Aussel}, {Ilbert},
  {Scoville}, {Surace}, {Frayer}, {Sheth}, {Helou}, {Brooke}, {Bhattacharya},
  {Yan}, {Kartaltepe}, {Barnes}, {Blain}, {Calzetti}, {Capak}, {Carilli},
  {Carollo}, {Comastri}, {Daddi}, {Ellis}, {Elvis}, {Fall}, {Franceschini},
  {Giavalisco}, {Hasinger}, {Impey}, {Koekemoer}, {Le F{\`e}vre}, {Lilly},
  {Liu}, {McCracken}, {Mobasher}, {Renzini}, {Rich}, {Schinnerer}, {Shopbell},
  {Taniguchi}, {Thompson}, {Urry}, \& {Williams}}]{Sanders2007}
{Sanders}, D.~B., {Salvato}, M., {Aussel}, H., {et~al.} 2007, \apjs, 172, 86,
  \dodoi{10.1086/517885}

\bibitem[{{Shipley} {et~al.}(2018){Shipley}, {Lange-Vagle}, {Marchesini},
  {Brammer}, {Ferrarese}, {Stefanon}, {Kado-Fong}, {Whitaker}, {Oesch},
  {Feinstein}, {Labb{\'e}}, {Lundgren}, {Martis}, {Muzzin}, {Nedkova},
  {Skelton}, \& {van der Wel}}]{Shipley:2018}
{Shipley}, H.~V., {Lange-Vagle}, D., {Marchesini}, D., {et~al.} 2018, \apjs,
  235, 14, \dodoi{10.3847/1538-4365/aaacce}

\bibitem[{{Silva} {et~al.}(1998){Silva}, {Granato}, {Bressan}, \&
  {Danese}}]{Silva1998}
{Silva}, L., {Granato}, G.~L., {Bressan}, A., \& {Danese}, L. 1998, \apj, 509,
  103, \dodoi{10.1086/306476}

\bibitem[{{Stefanon} {et~al.}(2015){Stefanon}, {Marchesini}, {Muzzin},
  {Brammer}, {Dunlop}, {Franx}, {Fynbo}, {Labb{\'e}}, {Milvang-Jensen}, \& {van
  Dokkum}}]{Stefanon2015}
{Stefanon}, M., {Marchesini}, D., {Muzzin}, A., {et~al.} 2015, \apj, 803, 11,
  \dodoi{10.1088/0004-637X/803/1/11}

\bibitem[{{Stone} {et~al.}(2024){Stone}, {Alberts}, {Rieke}, {Bunker}, {Lyu},
  {P{\'e}rez-Gonz{\'a}lez}, {Shivaei}, \& {Zhu}}]{Stone2024}
{Stone}, M.~A., {Alberts}, S., {Rieke}, G.~H., {et~al.} 2024, arXiv e-prints,
  arXiv:2405.18470, \dodoi{10.48550/arXiv.2405.18470}

\bibitem[{{Takagi} {et~al.}(2012){Takagi}, {Matsuhara}, {Goto}, {Hanami}, {Im},
  {Imai}, {Ishigaki}, {Lee}, {Lee}, {Malkan}, {Ohyama}, {Oyabu}, {Pearson},
  {Serjeant}, {Wada}, \& {White}}]{Takagi:2012}
{Takagi}, T., {Matsuhara}, H., {Goto}, T., {et~al.} 2012, \aap, 537, A24,
  \dodoi{10.1051/0004-6361/201117759}

\bibitem[{{Wang} {et~al.}(2024){Wang}, {Leja}, {Labb{\'e}}, {Bezanson},
  {Whitaker}, {Brammer}, {Furtak}, {Weaver}, {Price}, {Zitrin}, {Atek}, {Coe},
  {Cutler}, {Dayal}, {van Dokkum}, {Feldmann}, {Marchesini}, {Franx},
  {F{\"o}rster Schreiber}, {Fujimoto}, {Geha}, {Glazebrook}, {de Graaff},
  {Greene}, {Juneau}, {Kassin}, {Kriek}, {Khullar}, {Maseda}, {Mowla},
  {Muzzin}, {Nanayakkara}, {Nelson}, {Oesch}, {Pacifici}, {Pan}, {Papovich},
  {Setton}, {Shapley}, {Smit}, {Stefanon}, {Suess}, {Taylor}, \&
  {Williams}}]{Wang2024}
{Wang}, B., {Leja}, J., {Labb{\'e}}, I., {et~al.} 2024, \apjs, 270, 12,
  \dodoi{10.3847/1538-4365/ad0846}

\bibitem[{{Weaver} {et~al.}(2022{\natexlab{a}}){Weaver}, {Kauffmann}, {Ilbert},
  {McCracken}, {Moneti}, {Toft}, {Brammer}, {Shuntov}, {Davidzon}, {Hsieh},
  {Laigle}, {Anastasiou}, {Jespersen}, {Vinther}, {Capak}, {Casey},
  {McPartland}, {Milvang-Jensen}, {Mobasher}, {Sanders}, {Zalesky}, {Arnouts},
  {Aussel}, {Dunlop}, {Faisst}, {Franx}, {Furtak}, {Fynbo}, {Gould}, {Greve},
  {Gwyn}, {Kartaltepe}, {Kashino}, {Koekemoer}, {Kokorev}, {Le F{\`e}vre},
  {Lilly}, {Masters}, {Magdis}, {Mehta}, {Peng}, {Riechers}, {Salvato},
  {Sawicki}, {Scarlata}, {Scoville}, {Shirley}, {Silverman}, {Sneppen},
  {Smolc̆i{\'c}}, {Steinhardt}, {Stern}, {Tanaka}, {Taniguchi}, {Teplitz},
  {Vaccari}, {Wang}, \& {Zamorani}}]{Weaver2022}
{Weaver}, J.~R., {Kauffmann}, O.~B., {Ilbert}, O., {et~al.} 2022{\natexlab{a}},
  \apjs, 258, 11, \dodoi{10.3847/1538-4365/ac3078}

\bibitem[{{Weaver} {et~al.}(2022{\natexlab{b}}){Weaver}, {Kauffmann}, {Ilbert},
  {McCracken}, {Moneti}, {Toft}, {Brammer}, {Shuntov}, {Davidzon}, {Hsieh},
  {Laigle}, {Anastasiou}, {Jespersen}, {Vinther}, {Capak}, {Casey},
  {McPartland}, {Milvang-Jensen}, {Mobasher}, {Sanders}, {Zalesky}, {Arnouts},
  {Aussel}, {Dunlop}, {Faisst}, {Franx}, {Furtak}, {Fynbo}, {Gould}, {Greve},
  {Gwyn}, {Kartaltepe}, {Kashino}, {Koekemoer}, {Kokorev}, {Le Fevre}, {Lilly},
  {Masters}, {Magdis}, {Mehta}, {Peng}, {Riechers}, {Salvato}, {Sawicki},
  {Scarlata}, {Scoville}, {Shirley}, {Silverman}, {Sneppen}, {Smolcic},
  {Steinhardt}, {Stern}, {Tanaka}, {Taniguchi}, {Teplitz}, {Vaccari}, {Wang},
  \& {Zamorani}}]{Weaver2022_catalog}
---. 2022{\natexlab{b}}, VizieR Online Data Catalog, J/ApJS/258/11,
  \dodoi{10.26093/cds/vizier.22580011}

\bibitem[{{Williams} {et~al.}(2023){Williams}, {Tacchella}, {Maseda},
  {Robertson}, {Johnson}, {Willott}, {Eisenstein}, {Willmer}, {Ji}, {Hainline},
  {Helton}, {Alberts}, {Baum}, {Bhatawdekar}, {Boyett}, {Bunker}, {Carniani},
  {Charlot}, {Chevallard}, {Curtis-Lake}, {de Graaff}, {Egami}, {Franx},
  {Kumari}, {Maiolino}, {Nelson}, {Rieke}, {Sandles}, {Shivaei}, {Simmonds},
  {Smit}, {Suess}, {Sun}, {{\"U}bler}, \& {Witstok}}]{Williams2023}
{Williams}, C.~C., {Tacchella}, S., {Maseda}, M.~V., {et~al.} 2023, \apjs, 268,
  64, \dodoi{10.3847/1538-4365/acf130}

\bibitem[{{Wu} {et~al.}(2023){Wu}, {Ling}, {Goto}, {Kim}, {Hashimoto},
  {Kilerci}, {Lin}, {Wang}, {Uno}, {Ho}, \& {Hsiao}}]{wu:2023}
{Wu}, C. K.~W., {Ling}, C.-T., {Goto}, T., {et~al.} 2023, \mnras, 523, 5187,
  \dodoi{10.1093/mnras/stad1769}

\bibitem[{{Yang} {et~al.}(2023){Yang}, {Papovich}, {Bagley}, {Ferguson},
  {Finkelstein}, {Koekemoer}, {P{\'e}rez-Gonz{\'a}lez}, {Arrabal Haro},
  {Bisigello}, {Caputi}, {Cheng}, {Costantin}, {Dickinson}, {Fontana},
  {Gardner}, {Grazian}, {Grogin}, {Harish}, {Holwerda}, {Iani}, {Kartaltepe},
  {Kewley}, {Kirkpatrick}, {Kocevski}, {Kokorev}, {Lotz}, {Lucas},
  {Navarro-Carrera}, {Pentericci}, {Pirzkal}, {Ravindranath}, {Rinaldi},
  {Shen}, {Somerville}, {Trump}, {de la Vega}, {Wilkins}, \& {Yung}}]{Yang2023}
{Yang}, G., {Papovich}, C., {Bagley}, M.~B., {et~al.} 2023, \apjl, 956, L12,
  \dodoi{10.3847/2041-8213/acfaa0}

\bibitem[{{Young} {et~al.}(2023){Young}, {Pope}, {Sajina}, {Yan},
  {Gon{\c{c}}alves}, {Eleazer}, {Alberts}, {Armus}, {Bonato}, {Dale}, {Farrah},
  {Ferkinhoff}, {Hayward}, {McKinney}, {Murphy}, {Nesvadba}, {Ogle}, {Sajkov},
  \& {Veilleux}}]{Young2023}
{Young}, J., {Pope}, A., {Sajina}, A., {et~al.} 2023, \apjl, 958, L5,
  \dodoi{10.3847/2041-8213/ad07e1}

\bibitem[{{Zavala} {et~al.}(2021){Zavala}, {Casey}, {Manning}, {Aravena},
  {Bethermin}, {Caputi}, {Clements}, {Cunha}, {Drew}, {Finkelstein},
  {Fujimoto}, {Hayward}, {Hodge}, {Kartaltepe}, {Knudsen}, {Koekemoer}, {Long},
  {Magdis}, {Man}, {Popping}, {Sanders}, {Scoville}, {Sheth}, {Staguhn},
  {Toft}, {Treister}, {Vieira}, \& {Yun}}]{Zavala2021}
{Zavala}, J.~A., {Casey}, C.~M., {Manning}, S.~M., {et~al.} 2021, \apj, 909,
  165, \dodoi{10.3847/1538-4357/abdb27}

\end{thebibliography}
\bibliographystyle{aasjournal}

\end{document}